%% file: no_comments.tex
\documentclass[10pt,twocolumn]{article}
\usepackage[top=1.8cm, bottom=1.9cm, left=1.7cm, right=1.7cm]{geometry}
\usepackage{helvet}  %
\usepackage{courier}  %
\usepackage[hyphens]{url}  %
\usepackage{graphicx}  %
\usepackage{titling}
\usepackage{array,multirow,graphicx,adjustbox}  %
\usepackage{booktabs}                           %
\usepackage[utf8]{inputenc}
\usepackage[ruled,algosection,noend,linesnumbered]{algorithm2e}
\usepackage{float}
\usepackage{enumitem}
\usepackage{amsmath}
\usepackage{amsfonts}
\usepackage{xcolor}
\usepackage{csquotes} 
\usepackage{tabularx}
\usepackage{makecell}
\usepackage[sort&compress,numbers]{natbib}
\usepackage{pbox}
\usepackage{subfigure}
\usepackage{xspace}
\usepackage{paralist}
 \usepackage[small,bf]{caption}
 \usepackage[hang,flushmargin]{footmisc}

\renewenvironment{thebibliography}[1]{
  \begin{oldthebibliography}{#1}
    \setlength{\itemsep}{0.0em}
    \setlength{\parskip}{0.3em}
}
{
  \end{oldthebibliography}
}

\renewcommand{\footnoterule}{%
  \kern -3pt
  \hrule width 1in
  \kern 2pt
}

\makeatletter
\def\url@leostyle{%
  \@ifundefined{selectfont}{\def\UrlFont{}}%
  {\def\UrlFont{}}%
}
\makeatother
\usepackage[hyphenbreaks]{breakurl}

\definecolor{darkred}{RGB}{153,0,0}
\definecolor{darkblue}{RGB}{0,0,99}
\usepackage[colorlinks=true, linkcolor = darkred,   citecolor = darkred, urlcolor = darkblue]{hyperref}

\usepackage{times}
\renewcommand{\footnotesize}{\fontsize{8}{9}\selectfont}
\usepackage[compact]{titlesec}
\titlespacing*{\section}{0pt}{*4}{4pt}
\titlespacing{\subsection}{0pt}{*3}{3pt}
\titlespacing{\subsubsection}{0pt}{*2}{2pt}

\newcommand{\descr}[1]{\smallskip\noindent\textbf{#1}}

\newcommand{\dspol}{{{\selectfont /pol/}}\xspace}

\newif\ifcomment
\commenttrue

\ifcomment 
  \newcommand{\edc}[1]{\textbf{\em\color{red}#1}}
  \newcommand{\alex}[1]{\textit{\em\color{blue}#1}}
  \newcommand{\savvas}[1]{\textit{\em\color{orange} SZ: #1}}
  \newcommand{\jbnote}[1]{{\bf \textcolor{purple}{JB: #1}}}
  \newcommand{\edit}[1]{{\color{darkgreen}#1}}
  \newcommand{\red}[1]{{\color{red}#1}}
  \newcommand{\blue}[1]{{\color{blue}#1}}
\else
  \newcommand\edc[1]{}
  \newcommand\alex[1]{}
  \newcommand\savvas[1]{}
  \newcommand\jbnote[1]{}
  \newcommand\edit[1]{}
  \newcommand\red[1]{}
  \newcommand\blue[1]{}
\fi

\newif\ifshort
 \shortfalse %

\ifshort
  \newcommand{\isShort}{true}
\else
  \newcommand{\isShort}{false}
\fi
\newcommand{\shortVer}[1]{\ifthenelse{\equal{\isShort}{true}}{{#1}}{}}
\newcommand{\longVer}[1]{\ifthenelse{\equal{\isShort}{false}}{{#1}}{}}

\ifshort
\let\OLDthebibliography\thebibliography
\renewcommand\thebibliography[1]{
  \OLDthebibliography{#1}
  \setlength{\parskip}{0pt}
  \setlength{\itemsep}{1pt plus 0.2ex}
}
\else
\fi

\begin{document}

\sloppy

\title{\bf {\em ``And We Will Fight For Our Race!''} A Measurement Study of Genetic Testing Conversations on Reddit and 4chan\thanks{This is the full version of the paper appearing in the 14th AAAI Conference on Web and Social Media (ICWSM 2020). Please cite the ICWSM version.}\vspace*{-0.3cm}}

\author{Alexandros Mittos$^1$, Savvas Zannettou$^{2}$, Jeremy Blackburn$^3$, and Emiliano De Cristofaro$^1$\\[0.25ex]
\normalsize $^{1}$University College London $\;\;$ $^2$Cyprus University of Technology $\;\;$ $^3$Binghamton University\\
\normalsize \{a.mittos,e.decristofaro\}@ucl.ac.uk, sa.zannettou@edu.cut.ac.cy, jblackbu@binghamton.edu\vspace*{-0.3cm}}

\date{}

\maketitle

\begin{abstract}
Progress in genomics has enabled the emergence of a booming market for ``direct-to-consumer'' genetic testing. Nowadays, companies like 23andMe and AncestryDNA provide affordable health, genealogy, and ancestry reports, and have already tested tens of millions of customers. At the same time, alt- and far-right groups have also taken an interest in genetic testing, using them to attack minorities and prove their genetic ``purity.'' In this paper, we present a measurement study shedding light on how genetic testing is being discussed on Web communities in Reddit and 4chan. We collect 1.3M comments posted over 27 months on the two platforms, using a set of 280 keywords related to genetic testing. We then use NLP and computer vision tools to identify trends, themes, and topics of discussion. 

Our analysis shows that genetic testing attracts a lot of attention on Reddit and 4chan, with discussions often including highly toxic language expressed through hateful, racist, and misogynistic comments. In particular, on 4chan's politically incorrect board (/pol/), content from genetic testing conversations involves several alt-right personalities and openly antisemitic rhetoric, often conveyed through memes. Finally, we find that discussions build around user groups, from technology enthusiasts to communities promoting fringe political views. 
\end{abstract}

\section{Introduction}\label{sec:introduction}

Over the past decade, researchers have made tremendous progress toward understanding the human genome.
With increasingly low costs, millions of people can afford to learn about their genetic make-up, not only in diagnostic settings, but also to satisfy their curiosity about traits, wellness, or discover their ancestry and genealogy.

In this context, a few companies have successfully launched {\em direct-to-consumer (DTC)} genetic testing: individuals can purchase a kit (typically around \$100), mail it back with a saliva sample, and receive online reports after a few days. %
DTC companies offer a wide range of services, from %
romantic match-making\longVer{~\cite{dna2018romance} or identification of athletic skills~\cite{soccer2018genomics} }to reports of health risks\longVer{ (e.g, likelihood of developing Parkinson's)}, wellness\longVer{ (e.g., lactose intolerance)}, \longVer{carrier status (e.g., hereditary hearing loss), traits (e.g., eyes color), etc.}\shortVer{hereditary traits, etc.}
Popular products also include genetic {\em ancestry} tests, which promise a way to discover one's ancestral roots, 
building on patterns of genetic variations common in people from similar backgrounds~\cite{nih2018what}. 
\longVer{However, these are subject to limitations, e.g., results differ from provider to provider due to different control groups~\cite{ancestry2018science}. }%
AncestryDNA alone has tested more than 10M customers as of Jan 2019~\cite{ancestrydna2018about}.

Alas, increased popularity of self-administered genetic tests\longVer{, and in particular ancestry,} has also been accompanied by media reports of far-right groups using it to attack minorities or prove their genetic ``purity''\longVer{~\cite{business2016white,vice2016white} mirroring}\shortVer{~\cite{vice2016white}, prompting}
concerns of a new wave of scientific racism~\cite{newyork2018how}. %
\longVer{For example, white nationalists were recently taped chugging milk at gatherings to demonstrate the ability of white people to better digest lactose~\cite{newyork2018why}.}
Also, statements from Donald Trump led Senator Warren to publicly confirm her Native American ancestry via genetic testing, prompting heated debates on the matter~\cite{boston2018warren}. 

Interest in DTC genetic testing by right-wing communities comes at a time when racism, hate, and antisemitism on platforms like 4chan, Gab, and certain communities on Reddit is on the rise~\cite{Hine2017,Finkelstein2018}.
Thus, these trends are particularly worrying, also considering how technology has been disrupting society in previously unconsidered ways~\cite{gorodnichenko2018social}.
The fact that racist, misogynistic, and dangerous behavior festers and spreads on the Web at an unprecedented scale, eventually making its way into the real world, prompts the need for a thorough understanding of how genetic testing tools are being (mis)used in online discussions.
As genetics-based arguments for discrimination~\cite{watson}, and even genocide, have been made in the past, this should absolutely not be overlooked.
While other aspects of genetic testing have already been studied, e.g., how they affect one's perception of racial identity~\cite{Panofsky2017,Roth2018}, we are interested in the relation between genetic testing and online hate.
This is a topic that has not been thoroughly studied by the scientific community, despite, as discussed earlier, increasingly worrisome indications of far-right groups exploiting genetic testing for racist rhetoric.

With this motivation in mind, we identify and address the following research questions: 
(1) What is the overall prevalence of genetic testing discourse on social networks like Reddit and 4chan? 
(2) In what context do users discuss genetic testing? 
(3) Is genetic testing associated with far-right views, racist ideologies, hate speech, and/or white supremacy? 
(4) If so, in what context? Can we identify specific themes?

We compile and use a set of 280 keywords related to genetic testing to extract all available posts and comments from Reddit and 4chan. %
We collect 7K %
threads from the politically incorrect (\dspol) board of 4chan (consisting of 1.3M %
posts) from June 30, 2016 to March 13, 2018, and 77K %
comments from Reddit related to genetic testing from January 1, 2016 to March 31, 2018, and  analyze them along several axes to understand how genetic testing is being discussed online. 

We rely on natural language processing, computer vision, and machine learning tools, including: (1) Latent Dirichlet Allocation (LDA)\longVer{~\cite{blei2003latent}} to identify topics of discussion; 
(2) word embeddings\longVer{~\cite{mikolov2013distributed}} to uncover words used in a similar context across datasets; (3) Google's Perspective API~\cite{jigsaw2018perspective} to measure toxicity in texts; and (4) Perceptual Hashing to assess the imagery and memes shared in posts. 

\descr{Findings.} Overall, the {\em main} findings of our study include:\smallskip
\begin{compactenum}
\item Genetic testing is often discussed on \dspol and on subreddits associated with hateful, racist, and sexist content. These communities discuss genetic testing in a highly toxic manner, often suggesting its use to marginalize or even \emph{eliminate} minorities.\\[-1.5ex]
\item The analysis of images posted on \dspol highlights the recurrent presence of popular alt-right personalities and \enquote{popular} antisemitic memes along with genetic testing discussions.\\[-1.5ex]
\item Word embeddings analysis reveals that certain subreddits use ethnic terms in conjunction with genetic testing keywords in the same way as \dspol, which may be an indicator of 4chan's fringe ideologies spilling out on more mainstream Web communities.\\[-1.5ex] 
\item Genetic testing on Reddit is being discussed in a variety of contexts, e.g., dog breeds, crime evidence, and issues related to children (e.g., adoption, pregnancy). This indicates how mainstream genetic testing has become. \\[-1.5ex]
\item Reddit users are not uniformly interested in all aspects of genetic testing, rather, they form groups ranging from enthusiasts
 (e.g., those who are interested in or have undergone genetic testing), to people 
who use genetic keywords exclusively in subreddits that discuss fringe political views.
\end{compactenum}

\section{Related Work}\label{sec:related}

We now review relevant prior work. We report on three areas: 
1) studies on the societal effects of genetic testing;
2) measurement of hate speech on social networks; and 
3) exploratory studies on Reddit. 

\descr{Genetic Testing \& Society.}  
Panofsky and Donovan~\cite{Panofsky2017} analyze 70 discussion threads on the far-right website \url{stormfront.org}, where at least one user posted ancestry test results.  
They group posters based on whether they consider their results good and bad, and study how other Stormfront users react: if the posters receive \enquote{bad news,} they tend to question the validity of genetic genealogy science, trying to reinterpret their results to fit their views on races.
In follow-up work~\cite{panofsky2019genetic}, they also look at the relationship between citizen science and white nationalists' use of genetic testing, shedding light on how ``repair strategies'' combine anti-scientific attacks on the legitimacy of these tests and reinterpretations of them in terms of white nationalist histories. 
Mittos et al.~\cite{mittos201823andme} conduct an exploratory study of the Twitter discourse on genetic testing, examining 300K tweets related to genetic testing, and find that those who are interested in genetic testing appear to be tech-savvy and interested in digital health in general. 
They also find sporadic instances of people using genetic testing in a racist context, and others who express privacy concerns.

Chow et al.~\cite{chow2018warren} examine 2K tweets containing the keyword \enquote*{23andMe} spanning one week. 
They calculate their sentiment and find out that the positive tweets outnumber the negative, while users appear overall enthusiastic about the company's services.
Roth and Ivenmark~\cite{Roth2018} study how ancestry testing affects ethnic and racial identities, conducting 100 interviews with people who have white, black, Hispanic/Latino, Native American, and Asian ancestry. They find instances of consumers not accepting test results, and instead focus on estimates based on social appraisals and aspirations. Overall, they suggest genetic ancestry testing may reinforce race privilege. 
Clayton et al.~\cite{clayton2018systematic} conduct a meta-analysis of 53 studies involving 47K people around perceptions of genetic privacy, 
highlighting how survey questions are often phrased poorly, thus leading to possible misinterpretations of the results.
They also show that 
not enough attention was paid to influential factors, e.g., participants' sociocultural backgrounds. 
Finally, Couldry and Yu~\cite{couldry2018deconstructing} discuss how DTC genetic companies, such as 23andMe, influence the public toward sharing their genetic data by claiming that the abundance of data will improve people's lives in the long term, despite a body of work showing that genetic data cannot be securely anonymized~\cite{gymrek2013identifying,shringarpure2015privacy}.

Overall, most of the research in this area mostly relies on qualitative studies examining the societal effects of genetic testing~\cite{caulfield2012direct,darst2013perceptions,hann2017awareness,christofides2016company,nielsen2014perceptions}
and somewhat lacks quantitative large-scale measurements. 
To the best of our knowledge ours is the first large-scale, quantitative measurement study focusing on Reddit and 4chan. We examine trends, themes, and topics of discussion around genetic testing, and explore how communities related to the alt-right exploit genetic testing for sinister purposes. 

\descr{Online Hate.} Researchers have also studied hate speech on mainstream social networks like 
 \longVer{Twitter~\cite{silva2016analyzing,Davidson2017,Olteanu2018,Ribeiro2018}, Reddit~\cite{chandrasekharan2017bag,Olteanu2018}, Facebook~\cite{ben2016hate,del2017hate}}%
\shortVer{Twitter~\cite{silva2016analyzing,mondal2017measurement,Davidson2017,Olteanu2018}, Reddit~\cite{Olteanu2018}, Facebook~\cite{ben2016hate}}, YouTube~\cite{ottoni2018youtube}, and Instagram~\cite{Hosseinmardi2015}.
Closer to our work is research on fringe communities in 4chan and Reddit\longVer{~\cite{bernstein20114chan,Hine2017,Zannettou2017,Zannettou2018,Finkelstein2018}}. %
Specifically, Bernstein et al.~
\cite{bernstein20114chan} study 5M posts on the random (/b/) board to examine how anonymity and ephemerality work in 4chan, while Hine et al.~\cite{Hine2017} focus on \dspol, studying 8M posts collected over two and a half months. 
Their content analysis reveals that, %
while most URLs point to YouTube, a non-negligible amount link to right-wing websites. 
They also find evidence of organized ``raids'' against YouTube users, where users collectively post hateful comments on videos they disapprove.

Zannettou et al.~\cite{Zannettou2017} explore how mainstream and fringe online communities on Twitter, Reddit, and 4chan influence each other with respect to disinformation and hateful propaganda. 
Among other things, they find that racist memes are very common in \dspol and Gab, and that \dspol and the /r/The\_Donald subreddit are the most influential Web communities with respect to the dissemination of memes. 
In~\cite{Finkelstein2018}, authors study antisemitism on \dspol and Gab,
revealing that antisemitic content increases in those networks after major political events, such as the ``Unite the Right'' rally or the 2016 US elections. 
They also use word embeddings to identify terminology associated with antisemitic content.
Overall, while prior work identifies and/or measures hate on fringe platforms, we examine whether genetic testing, a seemingly harmless topic, is being discussed in a toxic manner.

\descr{Exploratory Studies on Reddit.} Another line of work has, similar to ours, performed quantitative studies using Reddit. 
De Choudhury and De~\cite{DeChoudhury2014Mental} look at Reddit conversations about mental health, aiming to understand language attributes of online self-disclosure and factors driving support in online posts. %
They show that 
users explicitly share personal information on their mental health, and use Reddit for self-expression, even for seeking diagnosis or treatment information.
Other studies analyze how users behave in specific subreddits.
Kasunic and Kaufman~\cite{Kasunic2018at} focus on \longVer{a specific subreddit called /r/RoastMe, where users post photos of themselves and invite others to ridicule them.
They find that the RoastMe community relies on a specific set of norms, such as highly valuing caustic comments but also being concerned about the potential psychological harm of the participants. }
Nobles et al.~\cite{Nobles2018std} study /r/STD to understand how users seek health information on sensitive and stigmatized topics, using 1.8K posts from 1.5K users. %
\longVer{They find that most posts crowd-source information about non-reportable STDs, focusing on treatment, symptoms, as well as aspects of social and emotional impact. 

}Another line of work has studied the /r/The\_Donald subreddit. 
Flores-Saviaga et al.~\cite{flores2018donald} analyze 16M comments spanning two years to examine the characteristics of political troll communities. 
They find that /r/The\_Donald subscribers spend energy educating their community on certain events and that they use various socio-technical tools to mobilize other subscribers. %
Finally, Mills~\cite{mills2018donald} compares /r/The\_Donald to /r/SandersForPresident, a subreddit broadly supporting the 2016 presidential candidate Bernie Sanders, exploring whether rapidly formed subreddits exhibit collective intelligence. 
Mills finds that these communities are very effective on pursuing their agendas and that Trump supporters more often tend to clash with other communities and Reddit administrators.

\section{Dataset}\label{sec:datasets}

In this section, we present the methodology used to obtain the datasets used in our study.

\descr{Genetic Testing Keywords.}
To extract relevant comments and posts we compile a list of 280 keywords related to genetic testing. 
First, we use the list of 268 DTC companies offering DNA tests over the Internet between 2011 and 2018 (e.g., 23andme, AncenstryDNA, Orig3n) obtained from~\cite{Phillips2018}.
We then add 12 more keywords: ancestry testing/test, genetic testing/test, genomic testing/test, genomics, genealogy testing/test, dna testing/test, and GEDMatch---an open data personal genomics database and genealogy website\longVer{~\cite{gedmatch}}. 

\descr{Reddit Dataset.} %
Reddit is a social news aggregation and discussion website, where users post content \longVer{(e.g., images, text, links)} which gets voted up or down by other users. 
Users can add comments to the posts, and comments can also be voted up or down and receive replies.
Top submissions appear on the front page, and top comments at the top of the post.
Content on Reddit is organized in communities created by users, {\em ``subreddits},'' which are usually associated with areas of interest (e.g., movies, sports, politics).
As of July 2019, Reddit has more than 330M monthly active users and 14B visits\longVer{, which makes it the fifth most visited site in the US~\cite{reddit2018}}. 

We gather all Reddit comments from January 1, 2016 to March 31, 2018 (2B comments in 473K subreddits) via the monthly releases from \url{pushshift.io}.\longVer{\footnote{\url{https://files.pushshift.io/reddit/}}}
We then use the 280 genetic testing keywords as search terms to extract all comments possibly related to genetic testing. 
This results in a dataset of 77K comments posted in 4.6K subreddits, as summarized in  Table~\ref{tab:datasets}. 
For comparison, we also obtain a set of 204K random comments unrelated to genetic testing. %

\begin{table}[t]
\centering
\setlength{\tabcolsep}{4pt}
\resizebox{0.99\columnwidth}{!}{%
\begin{tabular}{lrr|lrr}
  \toprule
  \textbf{\em Reddit}         & \textbf{Genetic}    & \textbf{Random}     & \textbf{\em 4chan}          & \textbf{Genetic}    & \textbf{Random}     \\ 
  & {\bf Testing} & & & {\bf Testing} \\
  \midrule
  Comments         & 77,184                      & 204,713             & Threads                              & 6,986                       & 19,530              \\
  Subreddits       & 3,734                       & 12,616              & Posts                                & 1,306,671                   & 760,691             \\ 
  Users            & 48,096                      & 165,127             & Posts/T (Mean)                       & 186.5                       & 37.9                \\ 
                   &                             &                     & Posts/T (Median)\hspace*{-0.5cm}     & 183                         & 5                   \\%
                   &                             &                     & Images                               & 338,540                       & 206,830             \\ \bottomrule 
\end{tabular}
}
\caption{Overview of the Reddit and 4chan datasets.}
\vspace{-0.2cm}

\label{tab:datasets}
\end{table}

\descr{4chan Dataset.} 
4chan is an imageboard website with virtually no moderation. 
An ``Original Poster'' (OP) creates a thread by posting an image and a message. 
Content is organized in subcommunities, called boards (as of January 2019, there are 70 of them), with various topics of interest (e.g., anime, sports, adult, politics, etc.).
Others can post in the OP's thread, with a message or an image. 
On 4chan, users do not need a registered account to post content. 
We focus on the politically incorrect board (\dspol), which has been shown to include a high volume of racist, xenophobic, and hateful content~\cite{Hine2017}.
We choose \dspol as we study how genetic testing is being discussed in communities that have been associated with alt-right ideologies. 

\begin{figure*}[t!]
  \center
  \includegraphics[width=0.7\textwidth]{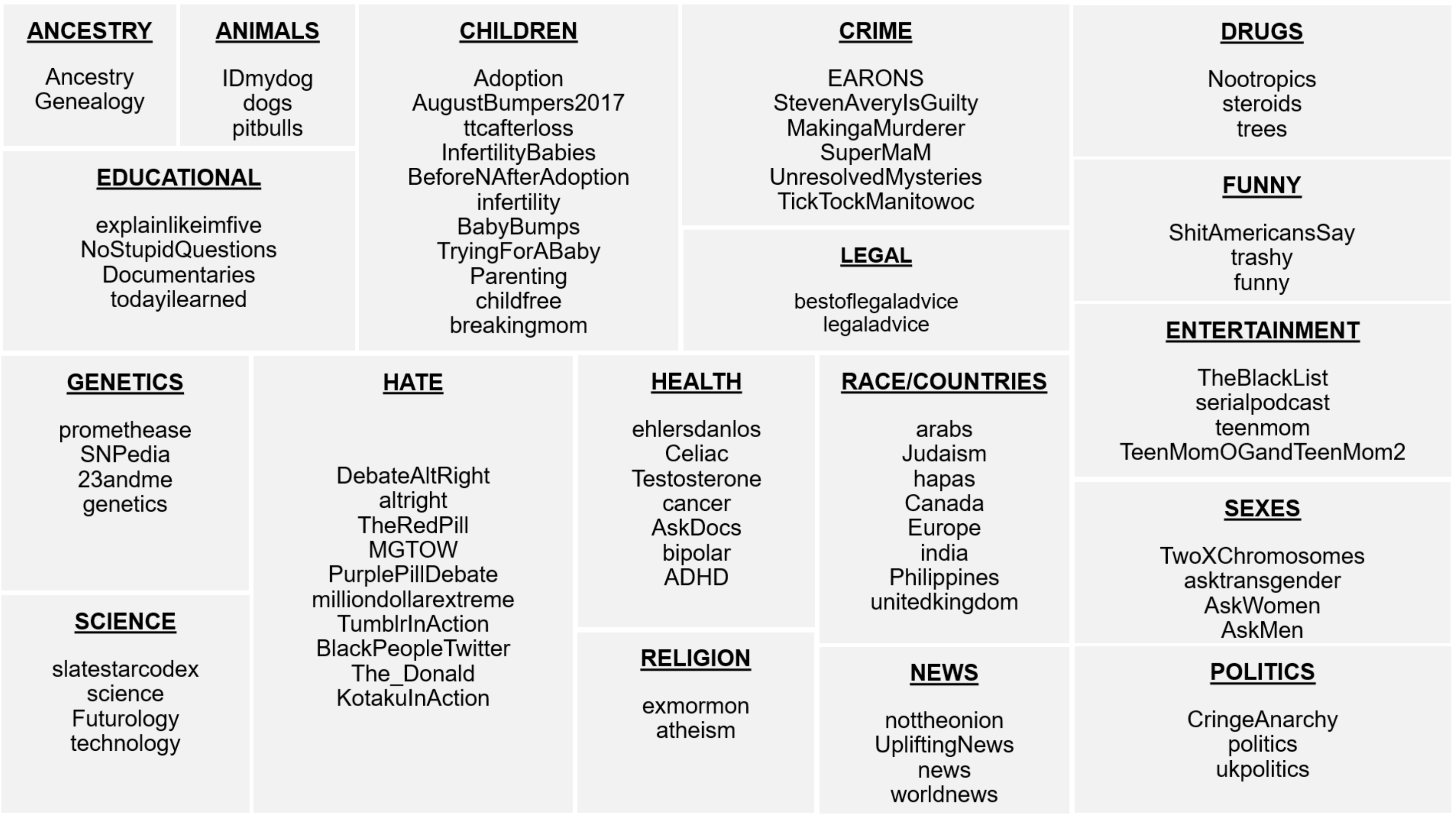}
  \caption{Subreddits with genetic testing related comments, grouped based on their thematic topics (excluding a generic `other' category).}
  \label{fig:topics_based_on_subreddits}
   \vspace{-0.3cm}
\end{figure*}

We collect 1.9M threads posted on \dspol from June 30, 2016 to March 13, 2018.
Once again, we use the 280 keywords as search terms on each thread: if we find a keyword anywhere in it, we get the \textit{whole thread}.
This is slightly different from what we do for Reddit.
On 4chan, each discussion is structured as a single-threaded entity where the OP submits an image on which other users respond. 
There is no official method of responding to a certain comment other than the original one, whereas, on Reddit a user may reply to a specific comment creating a new branch of answers. 
\longVer{Also, 4chan threads do not contain titles, thus, it is difficult to understand the context of a discussion without reading the whole thread. }%
In the end, we extract 6.9K threads containing 1.3M posts.
For comparison, we also get a random sample of 19K threads, 
with 760K posts.
The 4chan dataset is summarized in Table~\ref{tab:datasets}, where we report the mean and median number of posts per thread\longVer{, and the total number of images}.
\longVer{Note that, while the threads with genetic testing keywords have 338,540 images, later on we study only images shared in the {\em posts} containing those keywords (6,375).}

\descr{Ethics.} Our study was approved by the ethics committee at UCL. 
Also note that, as the content posted in 4chan is anonymous, we make no attempt to de-anonymize users.
Overall, we follow standard ethical guidelines~\cite{rivers2014ethical}.

\section{\hspace*{-0.1cm}Genetic Testing Discussions on Reddit}\label{sec:reddit}

In this section, we study genetic testing comments on Reddit. 
We start by identifying the subreddits with the highest number of comments related to genetic testing and thematically grouping them. 
Then, we use Google's Perspective API~\cite{jigsaw2018perspective}, a publicly available tool geared to identify toxic comments, to measure the toxicity of each group. 
We also use Latent Dirichlet Allocation (LDA) for basic topic modeling, aiming to extract the most prominent topics of discussion for each group. 
Finally, we examine comments that might be related to privacy concerns and perform a user analysis in terms of overlap across subreddits.

\subsection{Methodology}\label{sec:reddit_categories}

\descr{Subreddits selection \& grouping.}
We extract all the subreddits where genetic testing comments have been posted to,
but discard subreddits if they either have fewer than 1,000 comments overall or fewer than 100 comments with one of the keywords. %
This yields a list of 114 subreddits, which we list in Table~\ref{tab:most_common_subreddits} (see Appendix), along with the normalized number of genetic testing related comments. %

We group the subreddits into categories to study them based on (broad) discussion topics. 
We first turn to \url{redditlist.com}, a website reporting various subreddits metrics \longVer{(e.g., number of subscribers, growth, etc.) }and thematic tags, %
however, tags are available only for very popular subreddits\longVer{, and most of the subreddits in our list do not have them}.
Thus, we have two annotators browse the subreddits and assign up to five tags based on their thematic content. 
We then create a dictionary based on all the tags, and pick one tag which represents each subreddit best according to the annotators' judgment\longVer{ (the tag is also reported in Table~\ref{tab:most_common_subreddits})}.
Finally, we group them based on this tag, which leads to 18 categories plus a generic one, labeled as ``other'' (which includes 25 subreddits).
We report the subreddits in each category, except ``other,'' in Figure~\ref{fig:topics_based_on_subreddits}.

\longVer{Note that, while the content of most subreddits can be intuitively guessed from the name (e.g., /r/23andMe is about the company 23andMe), that is not always the case.
For instance, /r/AdviceAnimals is not about advice on animals, but on humans, and /r/trees is a subreddit about marijuana. 
Also, we opt to assign a separate \enquote*{Ancestry} category rather than \enquote*{Genetics}, since the former includes subreddits that do not necessarily deal with genetic testing.}

\descr{Prevalence of genetic testing comments.}
Unsurprisingly, the top five subreddits with most genetic testing comments are directly related to genetic testing/ancestry. 
Subreddits like /r/SNPedia or /r/Ancestry have a high fraction of comments with at least one genetic testing keyword; respectively, 10\% and 7\%.
We also find genetic testing to be relatively popular in subreddits about dog breed identification (/r/IDmydog, 1\%), children (/r/Adoption, 1\%), entertainment (/r/TheBlackList, 0.6\%), health (/r/ehlersdanlos, 0.7\%), and crime (e.g., /r/EARONS, 0.3\%).
By contrast, in the random dataset, only 6 out of 204K comments (0.003\%) include a genetic testing keyword.
Naturally, these percentages depict conservative lower bounds as:
1) comments can be replied to by other comments, thus creating different branches of discussion, and 
2) one can comment on a topic about genetic testing without using a keyword.
However, our approach provides ample data points for our analysis.
\longVer{
\begin{figure*}[t!]
  \center
  \subfigure[toxicity]{\includegraphics[width=0.33\textwidth,]{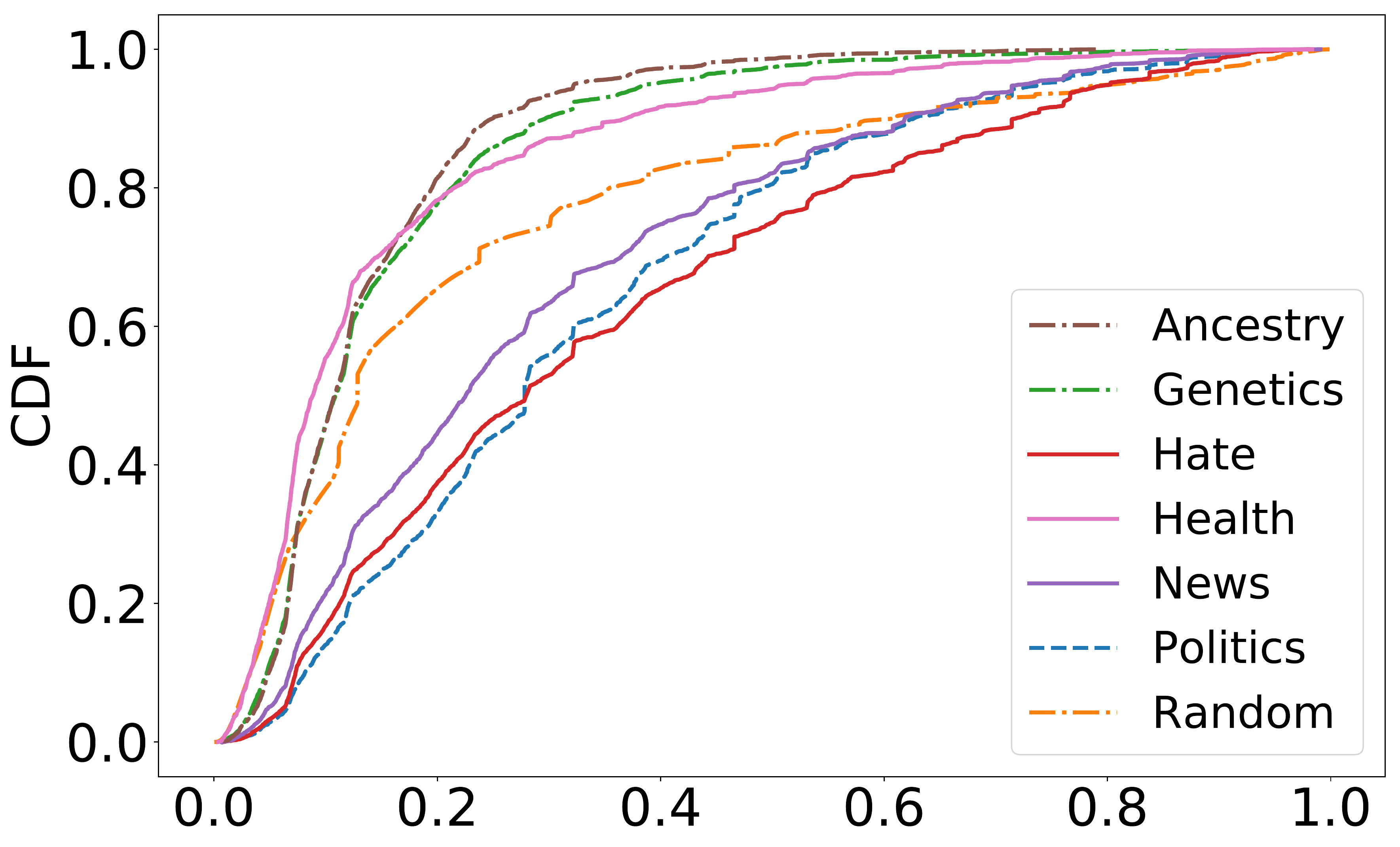}\label{fig:subreddits-toxicity}}
  \subfigure[severe toxicity]{\includegraphics[width=0.33\textwidth]{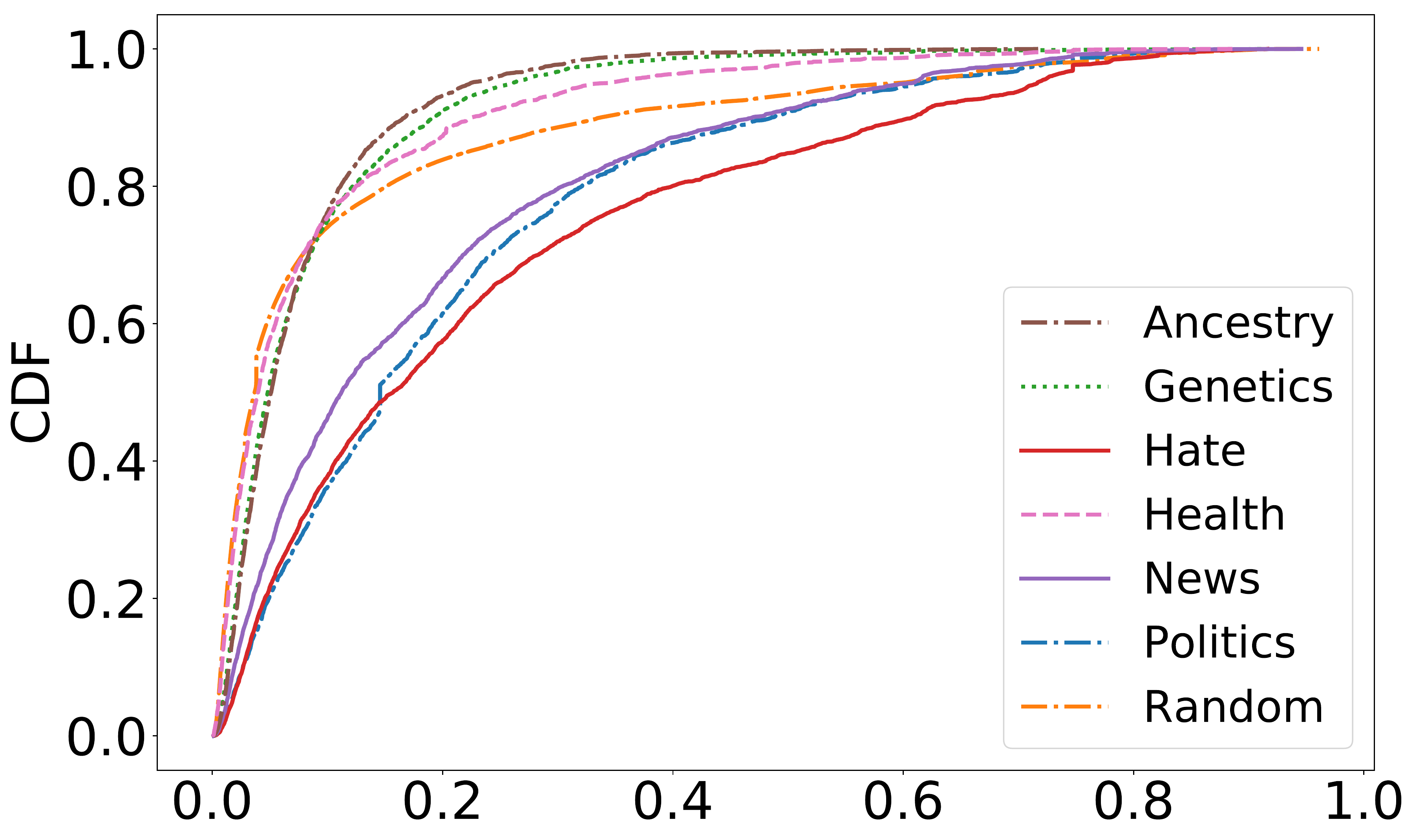}\label{fig:subreddits-severe-toxicity}}
  \subfigure[inflammatory]{\includegraphics[width=0.33\textwidth]{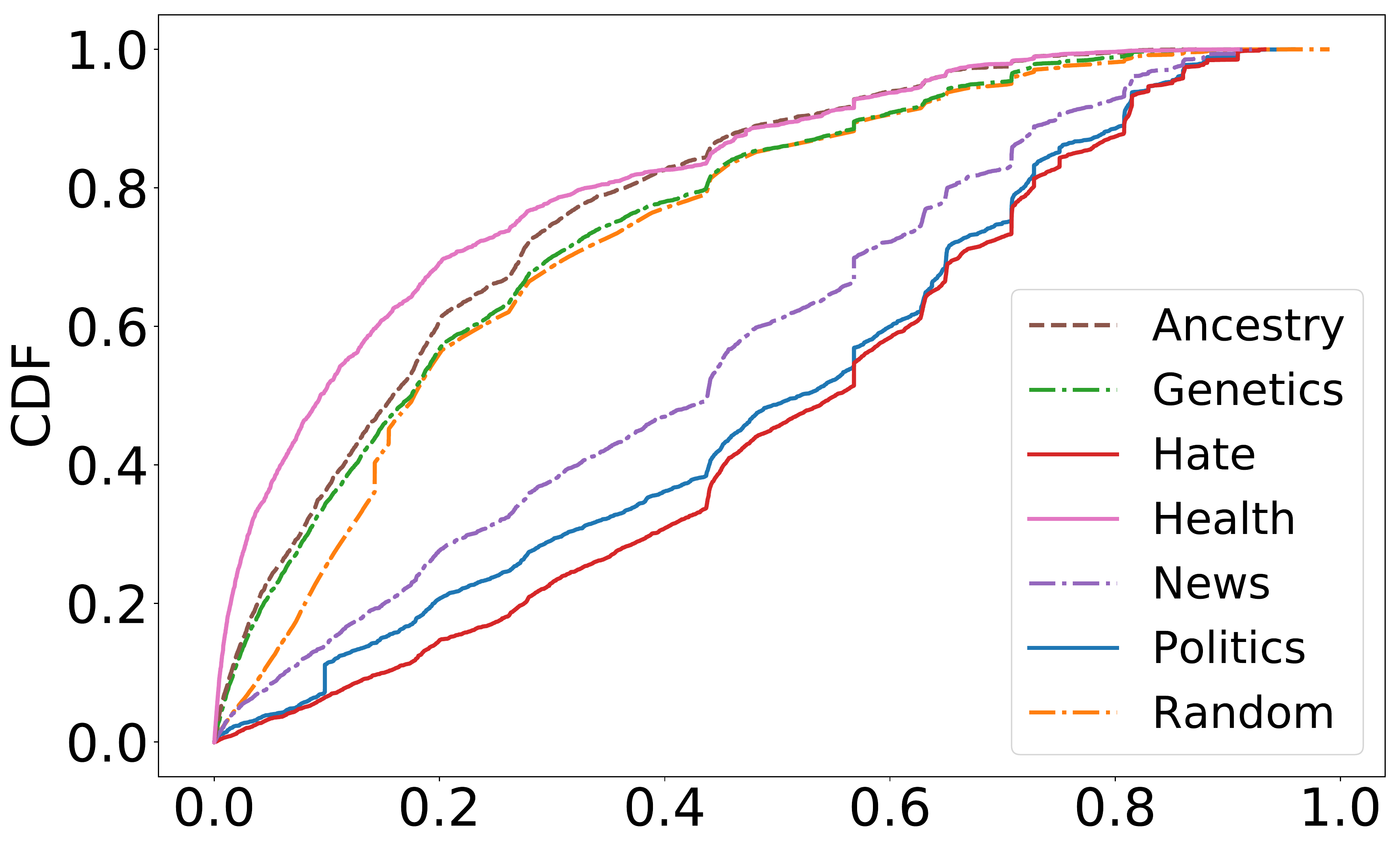}\label{fig:subreddits-inflammatory}}
  \caption{CDFs of Google's Perspective API toxicity metrics on the genetic testing comments for the three most and the three least toxic subreddit categories.}
  \label{fig:subreddits-perspective}
\end{figure*}
}

\shortVer{
\begin{figure*}[t!]
  \center
  \subfigure[toxicity]{\includegraphics[width=0.3\textwidth]{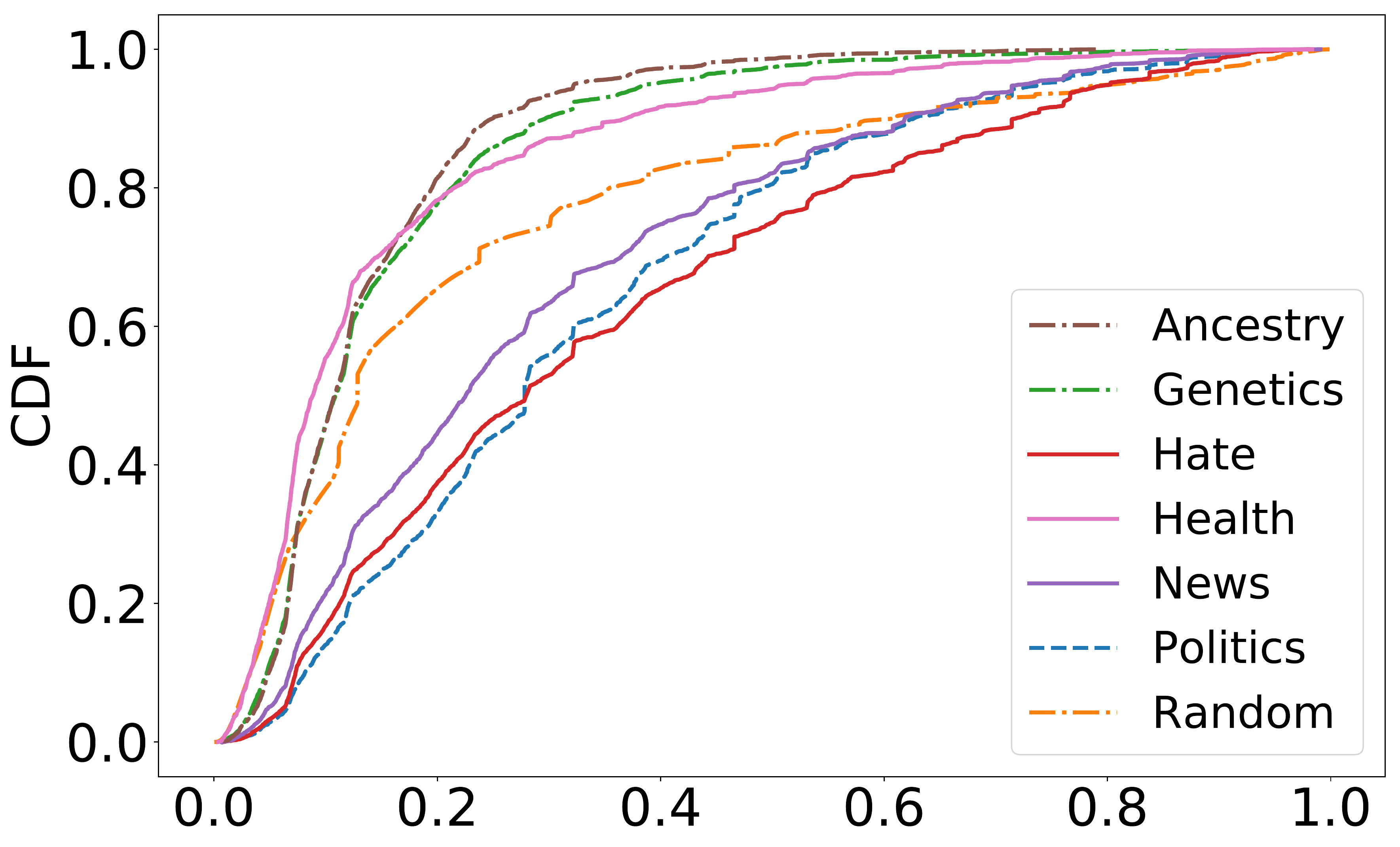}\label{fig:subreddits-toxicity}}
  \subfigure[severe toxicity]{\includegraphics[width=0.3\textwidth]{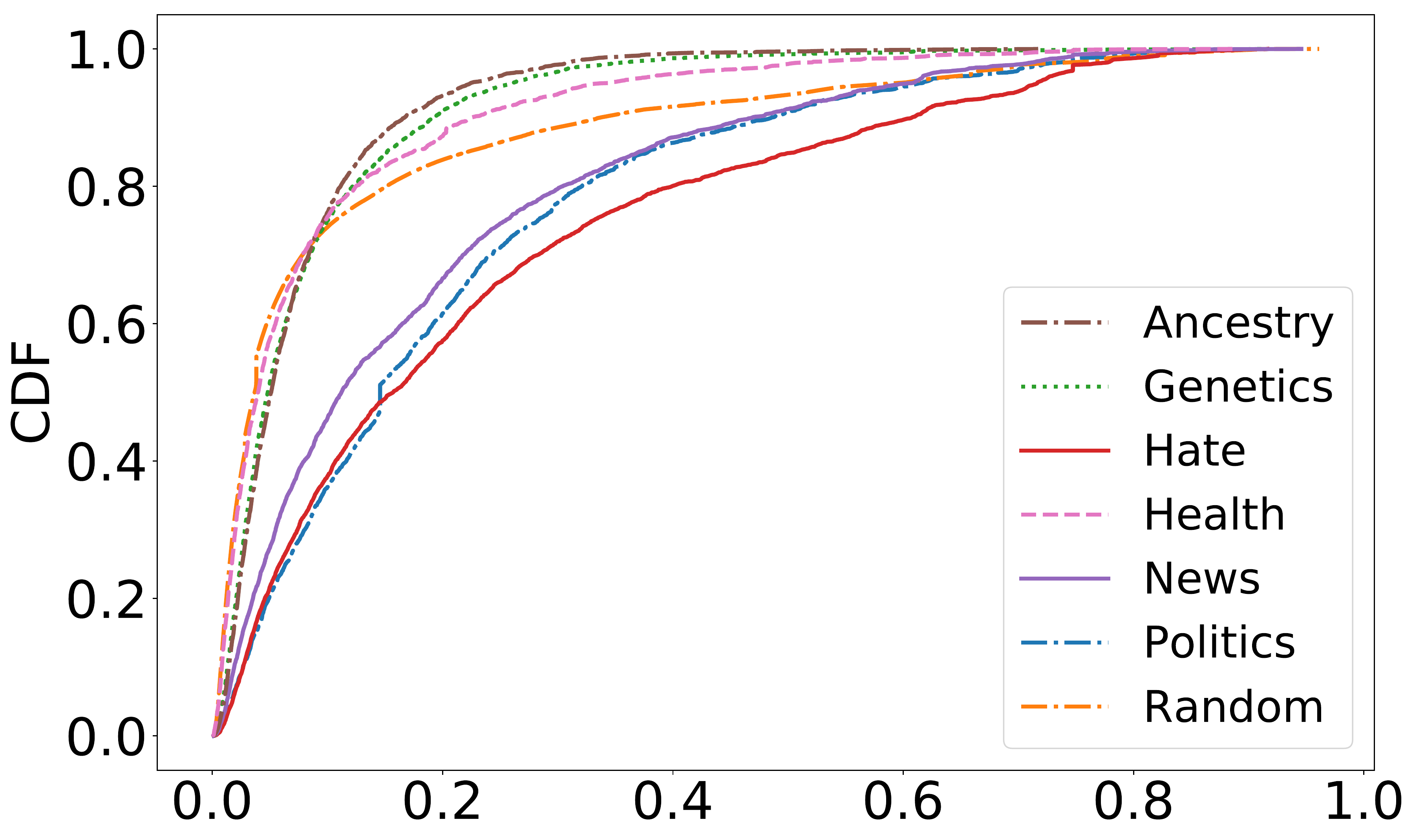}\label{fig:subreddits-severe-toxicity}}
  \subfigure[inflammatory]{\includegraphics[width=0.3\textwidth]{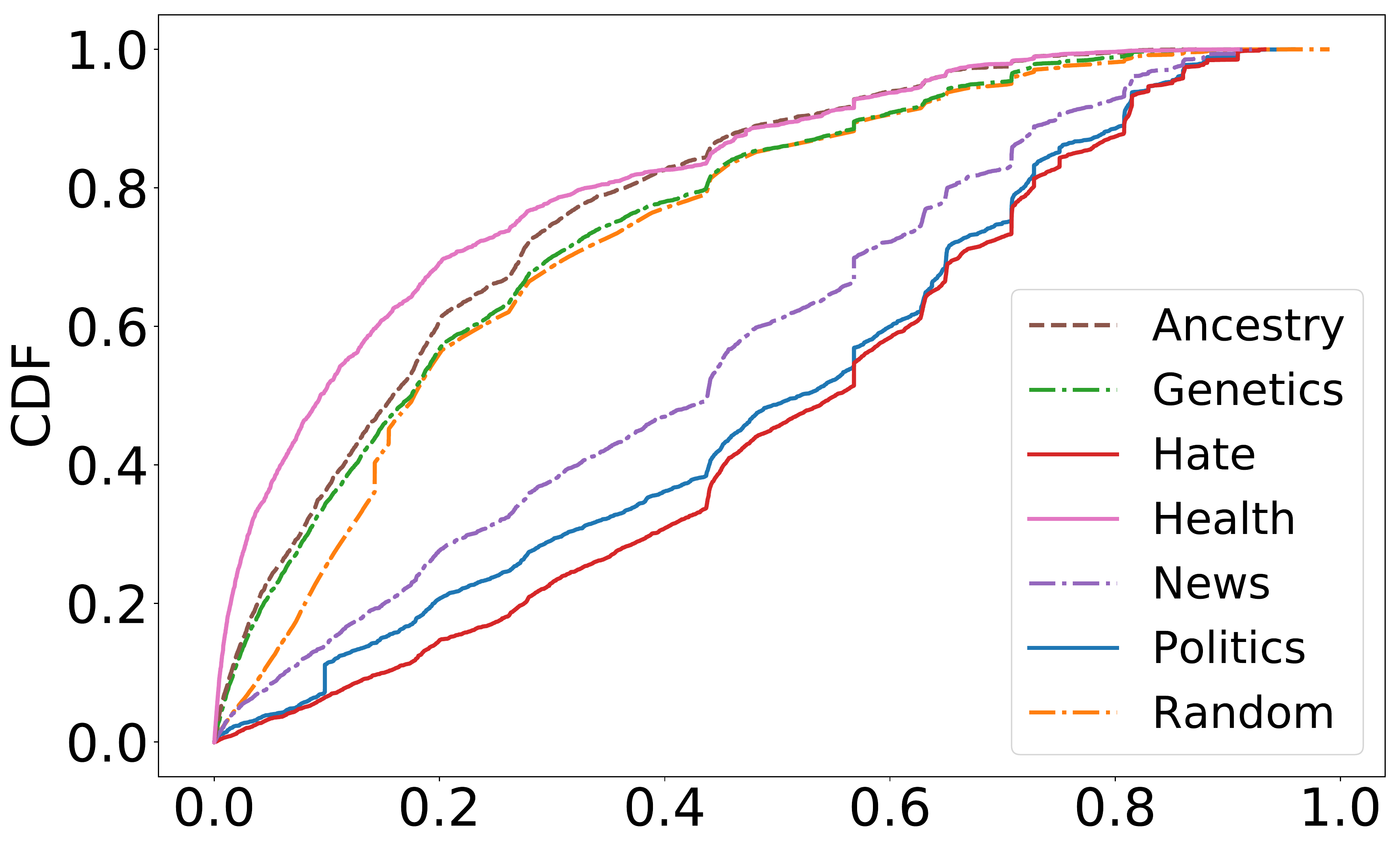}\label{fig:subreddits-inflammatory}}
   \vspace{-0.2cm}
  \caption{CDFs of Google's Perspective API toxicity on the genetic testing comments for the three most/least toxic subreddit categories.}
  \label{fig:subreddits-perspective}
\end{figure*}
}

\begin{table*}[t]
\centering
\setlength{\tabcolsep}{2pt}
\resizebox{\textwidth}{!}{
\begin{tabular}{rl}
\toprule
\textbf{Topic} & \multicolumn{1}{l}{\textbf{Category: Hate}}                                                                                                                    \\ \midrule
1   & dna (0.069), test (0.055), get (0.017), would (0.016), like (0.014), testing (0.013), know (0.012), one (0.011), think (0.009), take (0.008)                              \\
2   & child (0.037), men (0.023), women (0.022), father (0.019), woman (0.015), support (0.014), man (0.014), paternity (0.014), birth (0.011), get (0.008)                     \\
3   & white (0.034), people (0.021), african (0.016), black (0.015), european (0.013), race (0.013), ancestry (0.011), like (0.008), american (0.007), genetic (0.006)          \\
4   & jewish (0.028), native (0.017), american (0.015), israel (0.015), trump (0.013), clinton (0.010), jews (0.009), cherokee (0.007), citizenship (0.007), indian (0.007)     \\ 
5   & rep (0.027), dem (0.027), act (0.012), gay (0.007), body (0.007), gender (0.006), use (0.004), vote (0.004), proper (0.003), russia (0.003)                               \\
6   & testing (0.023), genetic (0.022), data (0.008), insurance (0.008), company (0.007), health (0.007), consent (0.007), paternity (0.006), companies (0.005), google (0.005) \\
7   & rape (0.021), women (0.012), lie (0.010), man (0.010), police (0.008), case (0.007), false (0.007), evidence (0.007), sex (0.006), point (0.005)                          \\
8   & genetic (0.016), human (0.006), even (0.006), testing (0.006), would (0.006), race (0.006), medical (0.006), differences (0.005), social (0.005), could (0.004)           \\
9   & youtube (0.010), talk (0.008), islamic (0.007), gedmatch (0.005), watch (0.005), working (0.005), video (0.005), dude (0.004), coast (0.004), saliva (0.004)              \\ 
10  & people (0.009), would (0.008), women (0.008), genetic (0.006), like (0.006), men (0.006), good (0.006), think (0.006), one (0.006), want (0.006)                          \\ \bottomrule
\end{tabular}
}
\vspace{-0.1cm}
\caption{LDA analysis of the Hate subreddits.}
\label{tab:lda_topics_hate}
\vspace{-0.2cm}
\end{table*}

\descr{Topics and toxicity.}\label{sec:subreddit_analysis} 
In the rest of this section, we analyze the 19 categories of subreddits in terms of the topics being discussed
as well as the toxicity of the comments therein,
using\longVer{, respectively,} LDA and Google's Perspective API~\cite{jigsaw2018perspective}.
The API returns three values between 0 and 1, pertaining to:
1) Toxicity, i.e., how rude, disrespectful, or unreasonable a comment is likely to be;
2) Severe Toxicity, which is similar to toxicity but only focuses on the ``most toxic'' comments; and 
3) Inflammatory, which focuses on texts intending to provoke or inflame. 
In Figure~\ref{fig:subreddits-perspective}, we plot the CDFs of the toxicity of the comments 
for the three most and the three least toxic subreddits (we also compare to the random dataset as a baseline).
We run two-sample Kolmogorov-Smirnov (KS) tests between the distribution of each category and the random dataset: in all cases, we reject the null hypothesis that they come from a common parent distribution ($p<0.01$).
We note that the two-sample KS test is non-parametric and thus robust in terms of different sample sizes.
While we acknowledge this might not be a perfect sampling, it is unlikely that any sampling method would result in perfectly balanced datasets. 
Also, recall that we are primarily interested in the overall comparison of content related (and unrelated) to genetic testing, thus this is appropriate for our purposes.
Overall, the comments originating from subreddits related to genetics, ancestry, and health are less toxic than a random baseline, while comments in news, politics, and ``hateful'' subreddits are remarkably more toxic. 

\descr{Remarks.}
We choose to use Google's Perspective to identify hateful content as other methods, e.g., hate speech detection libraries~\cite{Davidson2017}, are primarily trained on short texts with a limited number of training samples. Whereas, our datasets contain numerous lengthy comments which may span several thousand characters; thus, the Perspective API should perform better.
In the rest of the section, we report a few representative comments for each category based on our topic analysis -- i.e., interesting examples including words extracted as a topic.

\begin{table*}[t]
\centering
\setlength{\tabcolsep}{2pt}
\resizebox{\textwidth}{!}{
\begin{tabular}{rl}
\toprule
\textbf{Topic} & \multicolumn{1}{l}{\textbf{Category: Genetics}}                                                                                                                          \\ \midrule                                      
1   & dna (0.021), family (0.015), know (0.013), would (0.013), test (0.013), father (0.012), one (0.011), great (0.011), dad (0.009), mother (0.009)                                     \\ 
2   & european (0.023), ancestry (0.023), dna (0.017), african (0.015), results (0.014), people (0.014), native (0.013), american (0.012), eastern (0.011), german (0.009)                \\
3   & chromosome (0.031), haplogroup (0.031), ashkenazi (0.021), jewish (0.019), confidence (0.015), maternal (0.012), paternal (0.011), chromosomes (0.011), also (0.011), line (0.010)  \\
4   & genetic (0.021), testing (0.014), test (0.011), would (0.011), information (0.007), like (0.007), people (0.007), results (0.007), get (0.006), know (0.006)                        \\
5   & data (0.028), snps (0.020), one (0.013), snp (0.013), snpedia (0.011), gene (0.011), genome (0.010), raw (0.009), promethease (0.008), variant (0.008)                              \\
6   & blood (0.035), hair (0.023), eyes (0.018), type (0.017), cells (0.015), skin (0.015), blue (0.012), dark (0.011), brown (0.010), saliva (0.009)                                     \\
7   & asian (0.055), chinese (0.039), wegene (0.032), south (0.025), results (0.020), east (0.016), korean (0.014), japanese (0.014), southeast (0.013), customers (0.012)                \\
8   & sample (0.031), results (0.018), weeks (0.017), received (0.014), time (0.013), kit (0.013), samples (0.012), extraction (0.011), process (0.011), people (0.011)                   \\
9   & gedmatch (0.054), dna (0.044), data (0.033), ancestry (0.026), results (0.023), raw (0.020), upload (0.016), use (0.015), get (0.013), also (0.012)                                 \\
10  & ancestry (0.025), promethease (0.023), health (0.022), data (0.019), get (0.017), reports (0.017), report (0.014), new (0.011), results (0.011), ancestrydna (0.010)                \\ \bottomrule
\end{tabular}
}
\vspace{-0.1cm}
\caption{LDA analysis of the Genetics subreddits.}
\label{tab:lda_topics_genetics}
\vspace{-0.2cm}
\end{table*}

\begin{table*}[t]
\centering
\setlength{\tabcolsep}{2pt}
\resizebox{\textwidth}{!}{
\begin{tabular}{rl}
\toprule
\textbf{Topic} & \multicolumn{1}{l}{\textbf{Category: Ancestry}}                                                                                                            \\ \midrule
1 & match (0.029), dna (0.026), matches (0.025), one (0.016), cousins (0.014), shared (0.013), share (0.011), cousin (0.011), related (0.011), gedmatch (0.010)             \\
2 & dna (0.020), family (0.019), test (0.018), great (0.012), father (0.012), know (0.011), mom (0.011), would (0.011), mother (0.010), side (0.010)                        \\
3 & native (0.085), american (0.076), cherokee (0.018), ancestry (0.014), indian (0.011), nbsp (0.009), family (0.009), tribe (0.009), claim (0.008)                        \\
4 & dna (0.026), ancestry (0.018), results (0.011), irish (0.009), people (0.009), european (0.008), like (0.008), african (0.008), ethnicity (0.008), british (0.008)      \\
5 & william (0.019), youtube (0.016), watch (0.016), african (0.014), norwegian (0.013), sub (0.011), saharan (0.011), middle (0.009), census (0.008)                       \\
6 & dna (0.062), test (0.049), testing (0.020), father (0.020), would (0.019), autosomal (0.013), family (0.012), get (0.012), line (0.011), haplogroup (0.010)             \\
7 & ancestry (0.049), gedmatch (0.045), ftdna (0.028), dna (0.026), upload (0.024), results (0.024), test (0.023), matches (0.022), get (0.018), data (0.017)               \\
8 & jewish (0.031), european (0.023), asian (0.020), europe (0.018), east (0.017), eastern (0.017), italian (0.015), results (0.015), ancestry (0.014), ashkenazi (0.013)   \\
9 & dna (0.037), ancestry (0.018), ancestrydna (0.016), test (0.015), testing (0.013), data (0.010), tests (0.009), results (0.008), tree (0.008), information (0.007)      \\
10 & tree (0.029), find (0.018), family (0.017), people (0.016), trees (0.013), ancestry (0.012), see (0.012), records (0.012), matches (0.010), search (0.009)             \\ \bottomrule
\end{tabular}
}
\vspace{-0.1cm}
\caption{LDA analysis of the Ancestry subreddits.}
\label{tab:lda_topics_ancestry}
\vspace{-0.2cm}
\end{table*}
 
\subsection{Racism}

Remarkably, 10/114 subreddits in our sample are categorized as hateful as they are broadly associated with hateful content.
Some are clearly associated with the alt-right~\cite{newyorktimes2017alt}
(e.g., /r/altright, /r/DebateAltRight, and /r/The\_Donald), sexism, or racism.
For instance, /r/TheRedPill includes misogyny and toxic behavior towards women~\cite{guardian2016red}, while /r/MGTOW, Men Going Their Own Way, is a forum for men who reject romantic relationships with women\longVer{, and was identified as a supremacist group by the Southern Poverty Law Center~\cite{splc2017male}}. 
\longVer{Other subreddits in this group include /r/milliondollarextreme, an American sketch satire show associated with alt-right and antisemitism~\cite{atlantic2016million} which was banned in September 2018, 
as well as /r/KotakuInAction, which is associated with GamerGate-related toxicity~\cite{chatzakou2017measuring}. }%
Also, /r/BlackPeopleTwitter makes fun of tweets purporting to originate from African Americans.

With this in mind, we set to study the relation between genetic testing and racism on Reddit. 
Our Perspective API analysis (see Figure~\ref{fig:subreddits-perspective}) shows that the category related to hate is the most toxic, and some of the subreddits (e.g., /r/DebateAltRight, /r/altright) have among the highest number 
of comments including genetic testing keywords in this category of subreddits. %
In this context, the LDA modeling gives us insight on how these fringe communities discuss genetic testing; see Table~\ref{tab:lda_topics_hate}. 
Users often discuss their desire to get tested (e.g., dna, test, would, like, know), while others argue on issues related to paternity (e.g., paternity, father, support). 
Although we find similar topics in \longVer{genetics/ancestry and parenting}\shortVer{other} subreddits, here they are being expressed in a much more toxic/inflammatory manner, as shown by Figure~\ref{fig:subreddits-perspective}. %
For example, a user writes in /r/TheRedPill: \enquote{Would get a DNA test on those kids ASAP. I don't know why all men don't do them secretly as soon as the kids are born.}

Other topics are related to ancestry results (e.g., african, jewish, american, european) as well as race in general
(e.g., white, black, race), which are not as widely discussed in genetics/ancestry subreddits (see Tables~\ref{tab:lda_topics_genetics}, \ref{tab:lda_topics_ancestry}). 
Again, the conversations exhibit clear racist connotations; for example, a user writes in /r/DebateAltRight: \enquote{The Jews know who Jews are [...] It doesn't require genetic testing [...] We whites know who whites are. Non-whites know who whites are. Anyone with eyes knows who whites are. And we will fight for our race!}
\longVer{Finally, we find topics related to sexual crimes (e.g., \enquote*{rape}, \enquote*{women}, \enquote*{evidence}, \enquote*{sex}), homosexuality and gender (e.g., \enquote*{gay}, \enquote*{gender}), and insurance (e.g., \enquote*{insurance}, \enquote*{company}, \enquote*{health}).}

Overall, genetic testing is a relatively popular topic of discussion in subreddits associated with fringe political views. 
When looking at the comments with the highest toxicity, we find some disturbing content,
including instances of xenophobia (e.g., 
\longVer{\enquote{Can you be Alt-Right and have non-white friends?}, receiving the reply \enquote{No, as a member of the Alt-Right you have to DNA test all of your friends and if they're not 100\% White then you report them to your local Atomwaffen,} }%
\shortVer{\enquote{[...] as a member of the Alt-Right you have to DNA test all of your friends and if they're not 100\% White then you report them to your local Atomwaffen,} }%
 referring to a neo-nazi terrorist organization~\cite{atomwaffen2018}).
Some users explicitly advocate using genetic testing to eliminate groups of non-white ancestry (e.g., \enquote{You know with pre-implantation genetic testing we can breed out non-white ancestry fairly easily [...]}).

\subsection{Category Analysis}
Next, we select a few categories of subreddits and analyze them \shortVer{further}\longVer{via topic modeling and using toxicity metrics}, aiming to better understand how users perceive genetic testing in each context.
\longVer{To ease presentation, we only do so on interesting or unexpected categories.}

\descr{Genetics \& Ancestry.}
As mentioned, the subreddits with the highest ratio of genetic testing keywords\longVer{ (see top five subreddits in Table~\ref{tab:most_common_subreddits})} are directly related to genetic testing and ancestry. 
This is confirmed by LDA (see Tables~\ref{tab:lda_topics_genetics} and \ref{tab:lda_topics_ancestry}).
In fact, even in the genetics category, the discussion is dominated by ancestry (e.g., european, ashkenazi, african) and family (e.g., family, father, mother).
We also observe that the open personal genomics database and genealogy website, GEDmatch~\cite{gedmatch}, is one of the topics with the greatest weights (0.054); see Table~\ref{tab:lda_topics_genetics}. 
GEDmatch allows users to upload their genetic data obtained from DTC genetic testing companies to identify potential relatives who have also uploaded their data.
Interestingly, in December 2018, US police forces declared that GEDmatch helped them find suspects in 28 cold murder and rape cases~\cite{greytak2019genetic}.
Overall, as shown in Figure~\ref{fig:subreddits-perspective}, the subreddits about genetics and ancestry attract far less toxic comments than the random Reddit sample, and are the least toxic categories among the rest in our dataset. 
In particular, we observe extremely low levels of inflammatory content.

\begin{table*}[t]
\centering
\resizebox{\textwidth}{!}{
\begin{tabular}{rl}
\toprule
\textbf{Topic} & \multicolumn{1}{l}{\textbf{Category: Crime}}                                                                                                       \\ \midrule
1   & dna (0.041), would (0.020), testing (0.019), think (0.016), people (0.012), like (0.011), test (0.011), know (0.010), could (0.009), get (0.009)              \\
2   & blood (0.060), dna (0.043), testing (0.023), test (0.019), sample (0.013), vial (0.012), samples (0.012), tested (0.010), lab (0.009), tests (0.009)          \\
3   & found (0.016), murder (0.014), police (0.013), case (0.010), years (0.009), later (0.009), dna (0.008), man (0.007), went (0.007), convicted (0.006)          \\
4   & dna (0.054), test (0.020), evidence (0.019), testing (0.011), would (0.011), bullet (0.010), could (0.009), one (0.008), case (0.007), found (0.007)          \\
5   & one (0.011), would (0.007), control (0.007), lab (0.006), test (0.006), like (0.006), case (0.006), evidence (0.005), science (0.005), say (0.005)            \\
6   & evidence (0.023), avery (0.020), testing (0.016), dna (0.014), case (0.013), court (0.009), allen (0.008), trial (0.008), would (0.008), state (0.007)        \\
7   & father (0.031), family (0.023), mother (0.012), son (0.012), dad (0.011), related (0.011), adam (0.011), cousin (0.010), cousins (0.009), different (0.008)   \\
8   & said (0.019), fire (0.016), family (0.012), hobbs (0.008), brendan (0.007), barb (0.006), sketch (0.005), monday (0.005), richard (0.005), death (0.004)      \\
9   & avery (0.029), blood (0.017), would (0.017), evidence (0.017), found (0.015), key (0.011), garage (0.010), car (0.008), trailer (0.007), police (0.007)       \\
10  & bones (0.049), bone (0.035), remains (0.029), found (0.023), human (0.019), fragments (0.017), burn (0.016), pit (0.015), body (0.014), teresa (0.013)        \\ \bottomrule
\end{tabular}
}
\vspace{-0.1cm}
\caption{LDA analysis of the Crime subreddits.}
\label{tab:lda_topics_crime}
\vspace{-0.2cm}
\end{table*}

\begin{table*}[t]
\centering
\resizebox{\textwidth}{!}{
\begin{tabular}{rl}
\toprule
\textbf{Topic} & \multicolumn{1}{l}{\textbf{Category: Children}}                                                                                                        \\ \midrule
1   & testing (0.023), genetic (0.020), weeks (0.016), back (0.014), pregnancy (0.012), first (0.012), loss (0.010), results (0.010), pregnant (0.009), get (0.009)     \\  
2   & genetic (0.035), testing (0.017), child (0.017), children (0.014), people (0.012), health (0.012), would (0.011), kids (0.009), medical (0.008), life (0.008)     \\
3   & know (0.019), like (0.019), want (0.014), would (0.014), get (0.013), really (0.012), feel (0.011), time (0.010), think (0.009), even (0.009)                     \\
4   & child (0.050), dna (0.030), test (0.028), father (0.023), support (0.020), kid (0.016), dad (0.011), lawyer (0.011), paternity (0.010), get (0.009)               \\
5   & insurance (0.025), testing (0.013), get (0.013), genetic (0.012), doctor (0.012), labcorp (0.011), blood (0.009), pay (0.009), test (0.009), covered (0.008)      \\
6   & dna (0.030), test (0.020), family (0.019), parents (0.013), ancestry (0.012), also (0.011), birth (0.011), adoption (0.011), find (0.011), get (0.010)            \\
7   & weeks (0.034), genetic (0.029), scan (0.024), girl (0.022), testing (0.022), ultrasound (0.021), boy (0.016), baby (0.014), week (0.013), gender (0.012)          \\
8   & test (0.022), dna (0.016), name (0.012), back (0.012), got (0.008), came (0.008), said (0.008), little (0.008), son (0.007), chow (0.006)                         \\
9   & ivf (0.017), embryos (0.014), testing (0.011), one (0.010), genetic (0.010), pgs (0.010), dog (0.010), sperm (0.010), embryo (0.009), transfer (0.009)            \\
10  & genetic (0.032), testing (0.030), test (0.024), would (0.015), risk (0.012), baby (0.011), done (0.011), also (0.009), results (0.008), back (0.008)              \\ \bottomrule
\end{tabular}
}
 \vspace{-0.1cm}
\caption{LDA analysis of the Children subreddits.}
\label{tab:lda_topics_children}
\vspace{-0.2cm}
\end{table*}

\descr{Crime Investigations.} Genetic testing appears to be discussed frequently in subreddits falling in the crime category, e.g., /r/EARONS, the East Area Rapist/Original Night Stalker, a.k.a. the Golden State Killer~\cite{wired2018golden}.
We also find subreddits covering (often controversial) discussions about Steven Avery, who was wrongly convicted of sexual assault and attempted murder (this inspired Netflix's documentary Making a Murderer);
e.g., /r/StevenAveryIsGuilty seems to firmly believe Avery was justly convicted, while /r/TickTockManitowoc does not. %
The LDA analysis confirms how discussion in this category revolves around investigation, police, and evidence (e.g., blood, sample, vial, evidence); see Table~\ref{tab:lda_topics_crime}.

\longVer{A user writes: \enquote{Similarly, why didn't we get more impact out of the DNA test on the key? Specifically, one non-courtroom interviewee makes the point that TH DNA should have been all over the key, because she had owned it for many years. The fact that only SAs DNA was found seems to be evidence that it was in fact wiped/disinfected. Why wasn't this a bombshell to be used in court?}, while another says: \enquote{I am not convinced the DNA matched Teresa. I think they were probably random bones from a cadaver. Read about the DNA testing. It only matches in 7 of 15 locations.}}
The toxicity and inflammatory levels of the content of this category are similar to the random dataset, which, combined with the LDA results, suggest that genetic testing here is discussed for informational reasons. 

\descr{Parenting.}
Users also discuss genetic testing in the context of children, pregnancy, and parenting; e.g., in /r/Parenting, /r/Adoption, /r/TryingForABaby, /r/infertility. %
From the LDA analysis (see Table~\ref{tab:lda_topics_children}), we find that users often discuss topics related to the identity of the father or child support (e.g., father, support, lawyer), but also health and the characteristics of their child (e.g., ultrasound, gender, embryos). 
Once again, the subreddits in this category contain low levels of toxicity.

\descr{Animals.} 
Reddit users also use genetic testing keywords in subreddits related to animals, and more specifically those related to dogs. 
\shortVer{However, due to space limitations, we defer our analysis to the full version of the paper.}
\longVer{For instance, /r/IDmydog, which is the 8th ranked subreddit in terms of genetic testing comments, focuses on identifying dog breeds from pictures.
/r/dogs and /r/pitbulls focus on discussion about dogs and pitbulls respectively. %
This is also confirmed by LDA (e.g., breeds, terrier, mixed); see Table~\ref{tab:LDA_animals}. 
An interesting topic of discussion is related to dog breeds banned in certain countries~\cite{banned_breeds}, and how one can be identified through DNA testing.
For example, a user writes: \enquote{Why don't you get a DNA test done and see what he really is? If just by some chance he's not a banned breed, you can show that to your vet and get them to change the breed listed on record and then you can use that to show potential landlords if they say he looks like something that he's not.} %
Once again, this category has similar levels of toxicity and inflammatory content to the random dataset. }

\descr{Other categories.} 
Genetic testing is also discussed in educational contexts (e.g., /r/explainlikeimfive, /r/NoStupidQuestions), to learn about science (e.g., /r/science, /r/futurology), discuss their health (e.g., /r/celiac, /r/cancer), or in the context of drugs (/r/Nootropics, /r/steroids). 
Users also use words related to genetic testing in a legal context (/r/legaladvice), to discuss subjects related to their cultural background (e.g., /r/arabs, /r/judaism), as well as religion (e.g., /r/exmormon). 
Finally, we find genetic testing words in subreddits related to entertainment programs (e.g., /r/TheBlackList), comedy (e.g., /r/funny), and issues related to gender (e.g., /r/AskMen, /r/AskWomen).

\longVer{
\begin{table}[t]
\hspace*{-0.2cm}
\centering
\setlength{\tabcolsep}{1pt}
\resizebox{1.02\columnwidth}{!}{%
\begin{tabular}{lll}
\toprule
\textbf{Topic} & \textbf{Category: Animals}                                         \\ \midrule
1 & pit, breeds, breed, bull, amp, bulls, terrier, dogs, mixed, jpg                 \\
2 & dna, test, mix, like, dog, get, know, really, could, would                      \\
3 & dogs, genetic, still, testing, breed, thing, even, dog, issues, pedigree        \\
4 & health, genetic, breed, dogs, breeder, testing, breeding, breeders, dog, puppy  \\
5 & dog, breed, dna, banned, type, test, prove, one, court, pit                     \\ \bottomrule
\end{tabular}
}
\vspace{-0.1cm}
\caption{LDA analysis of the Animals subreddits.}
\label{tab:LDA_animals}
\vspace{-0.3cm}
\end{table}}

\subsection{Privacy Concerns}\label{sec:reddit_privacy}

We also examine comments where users discuss privacy concerns in the context of genetic testing. 
We do so motivated by the important privacy and security challenges prompted by the sensitive nature of genetic data~\cite{mittos2019systematizing}.
We extract comments which include both a genetic testing keyword and the word \enquote*{privacy} from our Reddit dataset, getting 560 comments (0.7\% of all comments).
Obviously, this set is a conservative sample, as it is possible for a user to discuss issues related to privacy without specifically mentioning the word \enquote*{privacy}. 
Then, we use LDA to identify the context in which users discuss issues related to privacy; see Table~\ref{tab:LDA_privacy}. 

\begin{figure*}[t!]
\vspace{-0.2cm}
  \center
\includegraphics[width=0.8\textwidth]{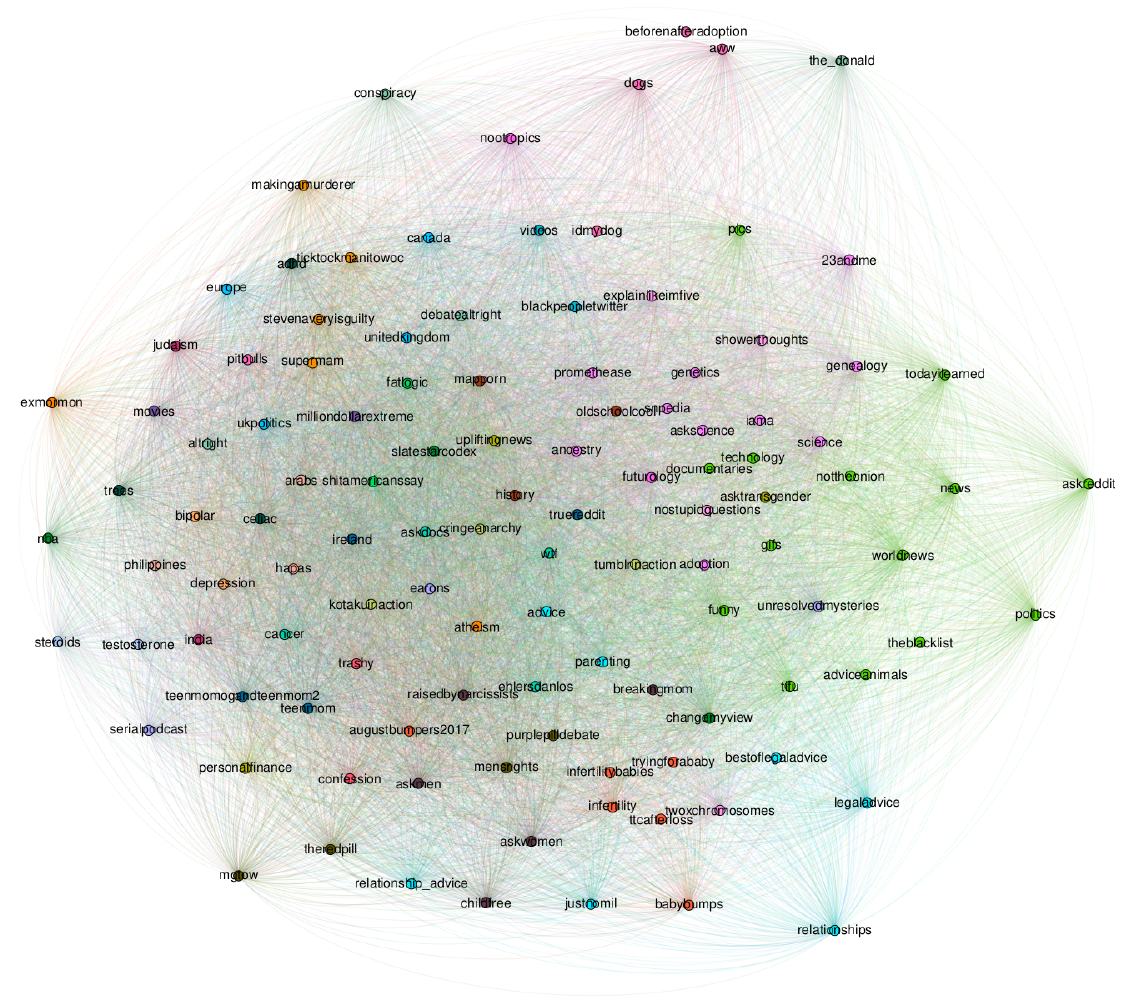}
  \vspace{-0.5cm}
  \caption{Graph depicting the Jaccard Index of the users whose comments include genetic testing keywords for each subreddit (see~\cite{fig3_interactive} for an interactive version).}
  \label{fig:jaccard_index_users}
\vspace{-0.2cm}
\end{figure*}

The most common subreddits in which privacy issues are being discussed are /r/genealogy, /r/news, and /r/23andMe. 
We find that Reddit users express privacy concerns on the use of genetic testing (e.g., dna, data, information, privacy, gender). 
Specifically, a topic of discussion is the potential misuse of genetic information by employers, while another topic focuses on paternity tests and whether children have the right to know their biological father. 
Finally, several users discuss the privacy issues stemming from a bill passed by the Republican Party on March 8, 2017, which allows companies to ask for their employees' genetic test results~\cite{begley2017house}. 

\begin{table}[t]
\hspace*{-0.2cm}
\centering
\setlength{\tabcolsep}{1pt}
\resizebox{1.02\columnwidth}{!}{%
\begin{tabular}{lll}
\toprule
\textbf{Topic} & \textbf{Category: Privacy}                                                           \\ \midrule
1 & dna, privacy, information, data, genetic, testing, ancestry, would, like, people                  \\
2 & would, child, test, people, one, privacy, father, think, know, right                              \\
3 & dna, genetic, health, information, employers, bill, wellness, sequencing, would, testing          \\
4 & dna, act, table, formatted, view, article, privacy, genetic, gender, testing                      \\
5 & cancer, breast, congress, house, trump, bill, genetic, republicans, act, laws                     \\ \bottomrule
\end{tabular}
}
\vspace{-0.1cm}
\caption{LDA analysis of the comments including \enquote*{privacy}.}
\label{tab:LDA_privacy}
\vspace{-0.3cm}
\end{table}

\subsection{User Analysis}

We also examine the overlap in users discussing genetic testing among all 114 subreddits in our sample. 
We do so to examine whether subreddits that have common interests have also similar user base.
For instance, we want to assess if users that post on /r/23andMe, also post on /r/ancestry.
To do so, we extract the set of users that posted in each subreddit and calculate the pairwise Jaccard Index scores between the set of users in each subreddit.
Next, we create a complete graph where nodes are the subreddits and edges are weighted by the Jaccard Index.
We then run the community detection algorithm in~\cite{blondel2008fast}, which provides a set of communities based on the graph's structure.

Figure~\ref{fig:jaccard_index_users} shows the resulting graph: nodes that have the same color are part of the same community.
The main observations are the following: 1) there are high Jaccard Index scores between the nodes in the same community, i.e., there is a substantial overlap of users that posted in all subreddits within the community.
2) Genetic testing subreddits (e.g., /r/genetics, /r/promethease, /r/ancestry, /r/23andMe)
are part of the same community (pink nodes) as scientific and education ones (e.g., /r/askscience, /r/science, /r/futurology), highlighting that ``enthusiasts'' are also active on scientific subreddits.
3) Subreddits associated with sexist content essentially share the same users (e.g., /r/MGTOW, /r/TheRedPill, /r/PurplePillDebate, lower left in olive green); also, users who discuss genetic testing in /r/The\_Donald are also active in other alt-right subreddits like /r/AltRight, /r/DebateAltRight (mint green nodes).

Additionally, we find communities with subreddits focused on the geopolitical aspects of genetic testing (see light blue nodes on the top left) like /r/europe, /r/canada, /unitedkingdom, and /r/ukpolitcs, as well as subreddits about  personal advice (light blue nodes on the bottom right) like /r/advice, /r/parenting, /r/legaladvice, /r/bestoflegaladvice.
Other communities are centered around conceiving children (e.g., /r/infertility, /r/tryingforababy, /r/babybumps, orange nodes on the bottom right side), crime investigation (e.g., /r/MakingaMurderer, /r/StevenAveryIsGuilty, orange nodes on the top left side), and animals (e.g., /r/dogs, /r/IDmydog, /r/pitbulls, pink nodes on top right side). 

\longVer{
Overall, Reddit users are not uniformly interested in every aspect of genetic testing, but rather specific communities focus on specific aspects thereof. 
For example, we find groups ranging from genetic testing enthusiasts, i.e., those who are interested in or have undergone genetic testing, to people who discuss genetic testing exclusively in subreddits with educational and scientific content, to those who use genetic testing terminology exclusively when discussing fringe political views.
}

\subsection{Take-Aways}

Our Reddit analysis shows that %
genetic testing %
is discussed in a variety of contexts which in itself is an indicator of how mainstream it has become. 
For instance, users discuss it in the context of issues related to their children, pets, or health, or to debate on their cultural heritage. %
More interestingly, they are not uniformly interested in every aspect of genetic testing, rather, they form {\em groups} ranging from genetic testing enthusiasts to individuals with fringe political views.
Thus, we observe a dichotomy in the type of users interested in genetic testing: 
some focus in typical uses of genetic testing, others discuss their use in worrying ways. 
Specifically, we find evidence of toxic language displaying clear racist connotations, and of groups of users using genetic testing to push racist agendas, e.g., to eliminate or marginalize minorities.
This is also particularly worrying since Reddit is a mainstream platform (5th most visited site in the US\longVer{~\cite{reddit2018}}).

\section{Genetic Testing Discussions on \dspol}\label{sec:4chan}

\longVer{We now study genetic testing comments on 4chan's politically incorrect board (\dspol). 
We first conduct a general characterization of the threads containing genetic testing keywords where we, similarly to the previous section, use Google's Perspective API to measure the toxicity of the contents and LDA modeling to extract the most prominent topics of discussion. 
Then, we use Perceptual Hashing~\cite{monga2006perceptual} and DBSCAN clustering~\cite{ester1996density} to study imagery and memes in the dataset.}
\shortVer{\vspace{0.1cm}}

\subsection{General Characterization}

\begin{figure}[t!]
  \center
\longVer{\includegraphics[width=0.75\columnwidth]{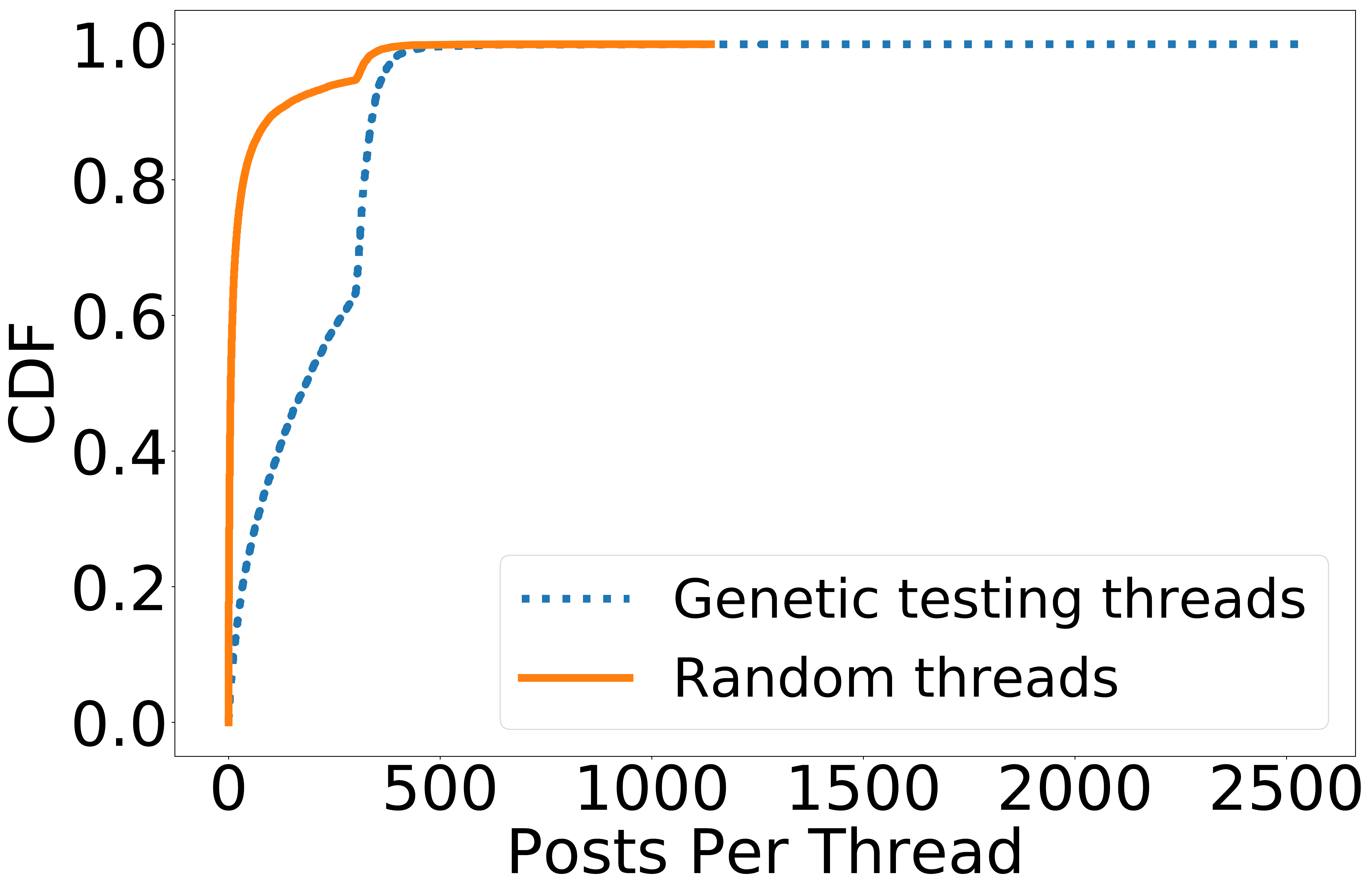}}
\shortVer{\includegraphics[width=0.55\columnwidth]{figures/CDF_4chan_posts_per_thread.pdf}}
 \vspace{-0.1cm}
  \caption{CDF comparing 4chan threads with genetic testing keywords and random threads in terms of number of posts.}
  \label{fig:4chan_cdf_posts_per_thread}
\vspace{-0.2cm}

\end{figure}

\descr{Thread Activity.} We begin by measuring the number of posts in threads where genetic testing keywords appear,
aiming to examine whether these threads attract more or less activity than ``usual.''
On \dspol, there is a limit on how many threads can simultaneously be active: whenever a new one is created, the one with the oldest last post is purged. 
There is also a ``bump'' limit that prevents a thread from never being purged.
As per~\cite{Hine2017}, the majority of threads attract only a few posts before being archived, while some---often covering controversial or popular topics---get many posts and possibly hit the bump limit.
\longVer{

}In Figure~\ref{fig:4chan_cdf_posts_per_thread}, we plot the CDF of the number of posts per thread, for both the genetic testing threads and our random sample. %
The former have an order of magnitude more posts than the latter (the median is 183 and 5 posts, respectively), %
which is an indicator that genetic testing is often discussed in long-lasting/interesting threads and may attract more attention by users.
We also run a two-sample Kolmogorov-Smirnov test on the distributions and we reject the null hypothesis that they come from a common parent distribution ($p<0.01$).

\descr{Toxicity \& Hate.} We then measure hate and toxicity in \dspol threads by computing:
1) percentage of hate words, 
and 2) toxicity/inflammatory levels.
For the former, we use a dictionary of hate words compiled by and available from \url{hatebase.org}, as used in~\cite{Hine2017}; for the latter, we again rely on the Perspective API.
However, we find no major differences between the genetic testing threads and the random sample---which is not surprising as \dspol is known for its high level of hate speech~\cite{Hine2017}---thus, we omit related plots to ease presentation.

\begin{table*}[t]
\centering
\setlength{\tabcolsep}{2pt}
\resizebox{\textwidth}{!}{
\begin{tabular}{rl}
\toprule
\textbf{Topic} & \multicolumn{1}{l}{\textbf{4chan}}                                                                                                           \\ \midrule
1 & ancestry (0.048), african (0.046), european (0.023), white (0.015), american (0.012), north (0.011), americans (0.010), population (0.008), south (0.008), europeans (0.008)  \\
2 & youtube (0.030), watch (0.028), jewish (0.020), king (0.013), company (0.010), lauren (0.010), tut (0.009), monkey (0.008), igenea (0.007), haplogroup (0.006)                \\
3 & ancient (0.023), modern (0.020), egyptians (0.015), egypt (0.012), years (0.009), national (0.008), egyptian (0.008), greeks (0.008), roman (0.007), saharan (0.007)          \\
4 & women (0.015), children (0.015), woman (0.011), men (0.010), man (0.009), genes (0.009), kids (0.009), child (0.008), two (0.008), birth (0.008)                              \\
5 & genetic (0.030), data (0.022), ancestrydna (0.014), information (0.014), health (0.013), company (0.012), testing (0.011), research (0.011), use (0.008), send (0.007)        \\
6 & back (0.022), got (0.021), european (0.020), family (0.020), german (0.013), took (0.012), irish (0.011), hair (0.011), came (0.011), eyes (0.010)                            \\
7 & dna (0.063), test (0.042), white (0.024), like (0.017), people (0.015), would (0.012), genetic (0.012), one (0.011), get (0.011), even (0.010)                                \\
8 & gedmatch (0.024), raw (0.014), creation (0.008), human (0.007), far (0.007), data (0.007), got (0.007), son (0.006), run (0.006), forum (0.006)                               \\
9 & screw (0.016), tweet (0.010), bill (0.010), tea (0.010), news (0.010), reddit (0.009), look (0.007), fda (0.005), search (0.005), guy (0.005)                                 \\
10 & companies (0.018), pay (0.016), child (0.015), order (0.015), racists (0.014), support (0.012), testing (0.011), adding (0.011), admit (0.011), law (0.011)                  \\ \bottomrule
\end{tabular}
}
\vspace{-0.1cm}
\caption{LDA analysis of \dspol.}
\label{tab:lda_topics_4chan}
\vspace{-0.2cm}
\end{table*}

\descr{Topic Modeling.} We also use LDA modeling to identify the most prominent topics of discussion; see Table~\ref{tab:lda_topics_4chan}.
Similar to Reddit, 4chan users use keywords suggesting their intention to get tested (e.g., would, get, dna, test). 
Several topics are related to ancestry, which is also among the words with the highest weights (0.048); for instance, users often discuss the ancestral background of the American population (e.g., american, african, european, white), others debate the cultural connection of modern humans to ancient civilizations (e.g., egyptians, greeks, roman), and the facial traits of modern europeans (e.g., german, irish, eyes, hair). 
Interestingly, another prominent topic of discussion is related to Lauren Southern (e.g., lauren, jewish, youtube), an Internet personality associated with the alt-right, whose popularity rose after being detained in Italy for trying to block a ship rescuing refugees~\cite{lauren2017former}. 
Other conversations likely relate to how genetic testing companies use their data (e.g., genetic, data, use, research), as well as legal issues related to child support (e.g., child, birth, support, law).

\subsection{Image Analysis} 
Next, we look at the images and memes that are shared in \dspol posts including genetic testing keywords.
We use the open source image analysis pipeline introduced in~\cite{Zannettou2018} which uses Perceptual Hashing~\cite{monga2006perceptual} and DBSCAN~\cite{ester1996density} to group together images that are visually similar.
We run the pipeline on the 6,375 images included in {\em posts} where at least one genetic testing keyword appears;
as discussed earlier, this is in contrast to the textual analysis where we look at whole threads.
\longVer{(Recall from Table~\ref{tab:datasets} that the total number of images in threads containing genetic testing keywords is 338,540.) }%
We obtain 215 clusters including 543 total images; the other 5,832 images are labeled as noise by the clustering algorithm and thus we discard them.
This high noise ratio mirrors findings in~\cite{Zannettou2018} and is likely due to 4chan users creating a lot of original content~\cite{Hine2017}. 
\longVer{Also, our dataset only includes a few thousand images, thus not a lot of images are visually similar.}

\begin{table}[t]
\centering
\resizebox{0.85\columnwidth}{!}{%
\begin{tabular}{@{}lrlr@{}}
\toprule
\textbf{Entity}          & \textbf{Clusters (\%)} & \textbf{Entity} & \textbf{Clusters(\%)}   \\ \midrule
\dspol                   & 15 (6.9\%)             & Video           & 3 (1.4\%)               \\
Lauren Southern          & 15 (6.9\%)             & Jewish people   & 3 (1.4\%)               \\
23andMe                  & 13 (6.0\%)             & Logo            & 3 (1.4\%)               \\
Pepe the Frog            & 9 (4.1\%)              & White           & 3 (1.4\%)               \\
United States of America & 8 (3.7\%)              & Shaun King      & 2 (0.9\%)               \\
Richard Spencer          & 5 (2.3\%)              & Screenshot      & 2 (0.9\%)               \\
Genetic                  & 4 (1.8\%)              & 4chan           & 2 (0.9\%)               \\
Meme                     & 4 (1.8\%)              & The Holocaust   & 2 (0.9\%)               \\
Europe                   & 3 (1.4\%)              & Race            & 2 (0.9\%)               \\
Greece                   & 3 (1.4\%)              & Adolf Hilter    & 2 (0.9\%)               \\ \bottomrule
\end{tabular}%
}
\vspace{-0.1cm}
\caption{Top 20 entities with the most clusters.}
\label{tbl:top_entities_images}
\vspace{-0.2cm}
\end{table}

We annotate each cluster using Google's Cloud Vision API.\footnote{\url{https://cloud.google.com/vision/}}
We calculate the medoid of each cluster (i.e., its ``representative'' image) %
following the methodology by~\cite{Zannettou2018}, and use that image to query the API.
This returns a set of meaningful entities, which are obtained by searching labeled images across the Web, along with their confidence scores. %
The exact methodology for extracting the entities is not known, however, upon manual examination, we can confirm that the API is indeed able to extract fine-grained entities.
For instance, given an image with Donald Trump, the API returns an entity called ``Donald Trump'' and not generic labels like ``man'' or ``politician.''

For each cluster, we extract the entity with the highest confidence score and analyze the top 20 entities, as reported in Table~\ref{tbl:top_entities_images}.
The most popular entries are \dspol itself and Lauren Southern with 6.9\% of all clusters. 
The latter is particularly interesting as it adds to the evidence that discussions about genetic testing frequently involve alt-right celebrities.
In fact, pictures of American white-supremacist Richard Spencer~\cite{spencer2016cnn} (6th most popular with 2.3\% of all clusters), and Carl Benjamin, a YouTuber known for his misogynistic involvement in the GamerGate controversy~\cite{vice2016carl}, are also popular. %

We also find several clusters related to: 1) 23andMe (6.0\%), e.g., screenshots of genetic testing results from 23andMe or images with the 23andMe logo, 2) memes including Pepe the Frog (4.1\%), a 4chan-popularized hate symbol~\cite{adl_pepe_frog}, and 3) geographic images related to, e.g., the US (3.7\%), Europe (1.4\%), or Greece (1.4\%).
The latter is likely mirroring discussions about the connection of modern humans to ancient civilizations; see topic 6 in Table~\ref{tab:lda_topics_4chan}.
We also find imagery related to the Jewish community (1.4\%), as well as the Holocaust (0.9\%) and Hitler (0.9\%), suggesting that, on 4chan, genetic testing terms and Nazi-related imagery are used together for the dissemination of hateful and antisemitic content. %

\begin{table*}[t]
\centering
\setlength{\tabcolsep}{2pt}
\resizebox{0.94\textwidth}{!}{
\begin{tabular}{rl}
\toprule
\textbf{Topic} & \multicolumn{1}{l}{\textbf{Entity: 23andMe}}                                                                                               \\ \midrule
1 & dna (0.050), ancestry (0.035), tests (0.024), results (0.018), one (0.018), percent (0.016), african (0.016), got (0.014), would (0.014), could (0.014) \\
2 & could (0.030), jewish (0.030), even (0.023), pol (0.023), people (0.023), also (0.023), company (0.016), test (0.016), results (0.016), markers (0.016) \\
3 & white (0.039), genetic (0.034), test (0.034), heritage (0.022), european (0.022), dna (0.018), jew (0.018), like (0.018), nigger (0.014), still (0.014) \\ \bottomrule
\end{tabular}
}
\vspace{-0.1cm}
\caption{LDA analysis of the texts in the \dspol posts with imagery annotated as \enquote*{23andMe}.}
\label{tab:lda_topics_entity_23andme}
\vspace{-0.2cm}
\end{table*}

\begin{table*}[t]
\centering
\setlength{\tabcolsep}{2pt}
\resizebox{\textwidth}{!}{
\begin{tabular}{rl}
\toprule
\textbf{Topic} & \multicolumn{1}{l}{\textbf{Entity: United Stated of America}}                                                                                          \\ \midrule
1 & white (0.044), ancestry (0.038), americans (0.031), self (0.028), african (0.021), european (0.018), even (0.018), whites (0.018), race (0.018), american (0.018)   \\
2 & white (0.039), roman (0.024), people (0.021), whites (0.018), full (0.018), empire (0.016), citizenship (0.016), held (0.016), admixture (0.016), like (0.016)      \\
3 & sargon (0.042), get (0.037), spencer (0.032), enoch (0.032), like (0.027), anyone (0.027), think (0.022), say (0.017), would (0.017), even (0.017)                  \\ \bottomrule
\end{tabular}
}
\vspace{-0.1cm}
\caption{LDA analysis of the texts in the \dspol posts with imagery annotated as \enquote*{United Stated of America}.}
\label{tab:lda_topics_entity_usa}
\end{table*}

We also examine the entities in Table~\ref{tbl:top_entities_images} more closely to shed light on the context in which images are being discussed.
Specifically, we extract text from the posts appearing alongside the images and use LDA modeling on the posts of each entity separately. %
We set LDA to produce only three topics per entity given the limited number of posts per entity. %
Among other things, we find that posts containing images related to 23andMe (see Table~\ref{tab:lda_topics_entity_23andme}) actually include discussions with racial connotations; for instance, whether test results show signs of African ancestry (e.g., ancestry, percent, african), or whether people with Jewish heritage are behind the company (e.g., jewish, company, results). 
For example, a user writes: \enquote{Can a genetics company founded by a Jew be trusted?}
Similarly, posts with images annotated as United States of America (see Table~\ref{tab:lda_topics_entity_usa}) reveal discussions on the ancestral background of the American population (e.g., americans, ancestry, african, whites). 
\longVer{A user writes: \enquote{Less than 5\% of White Americans have even negligible amounts of African DNA}. }

\longVer{
\begin{figure}[t!]
\centering
\includegraphics[width=0.67\columnwidth]{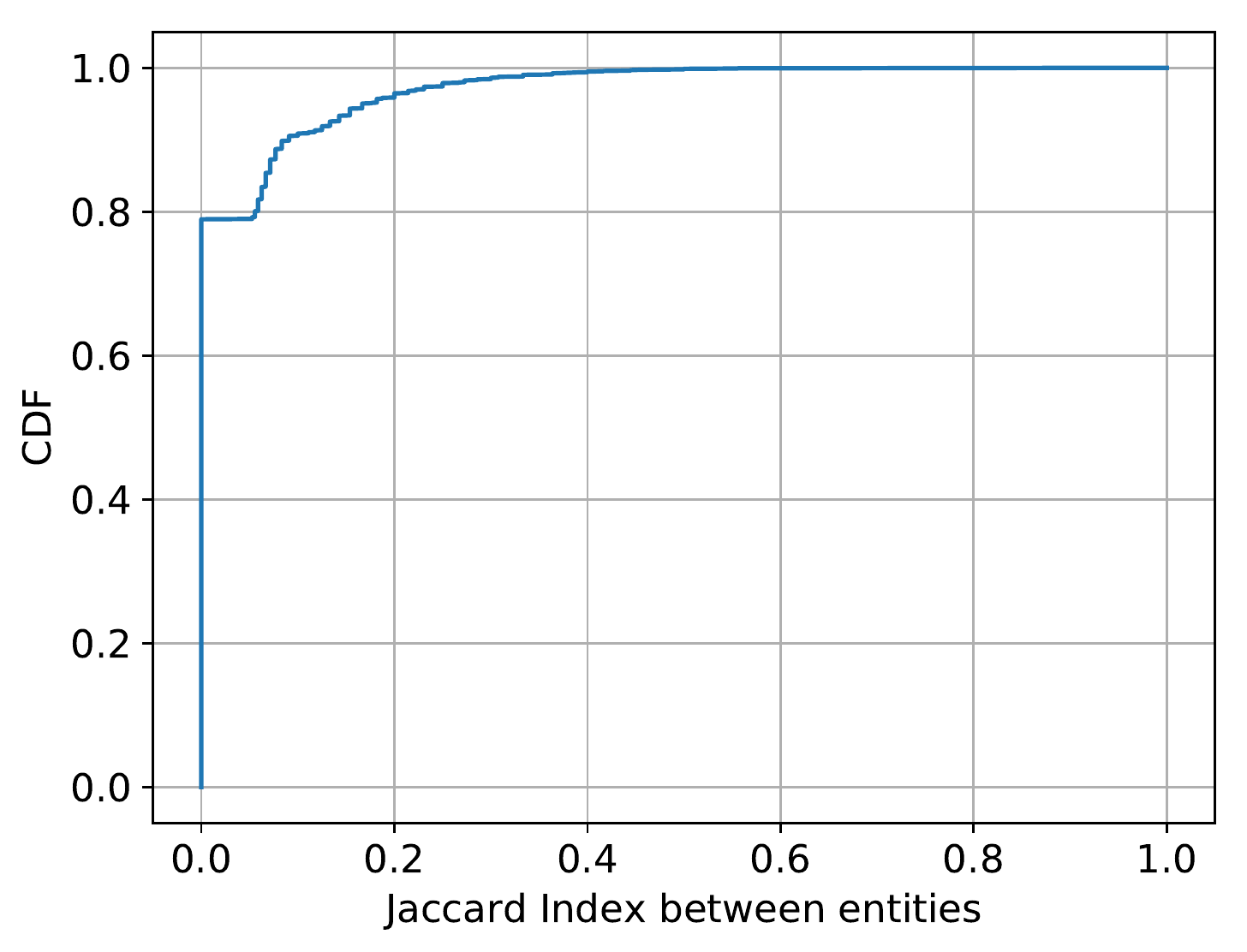}
\caption{CDF of the clusters' Jaccard Index scores using the set of entities returned by the Cloud Vision API.}
\label{fig:cdf_jaccard_scores}
\vspace{-0.2cm}
\end{figure}
}

\begin{figure*}[t!]
  \centering
\includegraphics[width=0.865\textwidth]{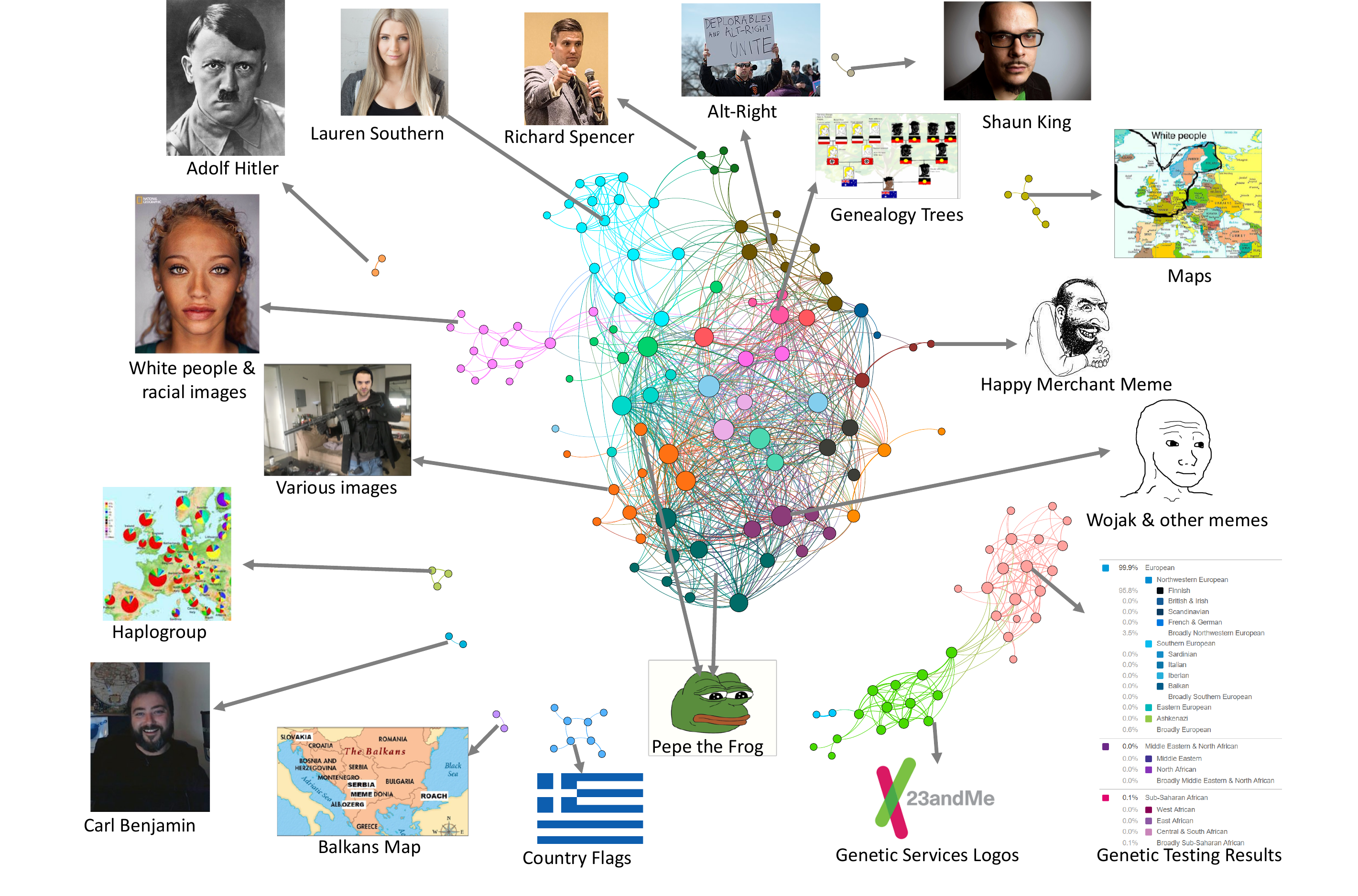}
   \vspace{-0.4cm}
  \caption{Visualization of the image clusters with manual annotation (see~\cite{fig6_interactive} for an interactive version).}
  \label{fig:clusters_islands}
\vspace{-0.2cm}
\end{figure*}

\descr{Cluster visualization.} Finally, we provide a visualization of the clusters in Figure~\ref{fig:clusters_islands}. 
Nodes in the graph represent clusters, while edges represent the Jaccard Index between clusters (as per the entities returned by the Cloud Vision API).\longVer{ To ease presentation, we only report edges where the Jaccard Index is greater than a threshold.
To select this threshold, we plot the CDF of all the Jaccard Index scores in Figure~\ref{fig:cdf_jaccard_scores}, which shows that  80\% of clusters are completely disjoint (Jaccard Index equal to 0).
As a result, we decide to select a $0.2$ threshold.}
\shortVer{To ease presentation, we only consider edges where the Jaccard Index is greater than 0.2, a threshold we select after inspecting the distribution of all the Jaccard Index scores.}
This corresponds to selecting 4.1\% of the edges with the highest Jaccard Index, allowing us to understand the {\em main} connections between clusters. 

Then, we perform community detection, using the approach presented in~\cite{blondel2008fast}. %
This considers the structure of the graph and decomposes it into a set of communities, where each community includes a set of highly inter-connected nodes. 
The resulting graph is presented in Figure~\ref{fig:clusters_islands}, with each color representing a different community.
For each community, we have manually inspected the images in the clusters and added a high-level description as well as a representative image.

The figure highlights the presence of two tightly-knit communities (bottom right): the green community includes images with logos of genetic testing companies, while the light red community covers images with screenshots of genetic testing results.
We also find communities with images related to Haplogroups and Genealogy Trees,
as well as others related to the alt-right (top of the graph).
In fact, a few communities exhibit clear racial connotations (pink), e.g., a cluster including an image from National Geographic predicting how the average American woman will look like in 2050~\cite{american_woman_2050}, which, unsurprisingly, attracted numerous posts on 4chan.
Finally, a few communities are related to hateful memes like Pepe the Frog and the Happy Merchant, a caricature of a manipulative Jew used on 4chan in racist contexts~\cite{Finkelstein2018}.

\subsection{Takeaways}

Overall, we find that genetic testing is a rather popular topic of discussion in 4chan's \dspol, %
often appearing in long/active threads.
Also, genetic testing topics are often accompanied by images and memes with clear racial or hateful connotations. 
While the presence of highly toxic content in \dspol is unsurprising, 
the specific content which accompanies threads related to genetic testing is very worrying. 
We find imagery with prominent figures of the alt-right movement (e.g., Lauren Southern, Richard Spencer), antisemitic memes (e.g., Pepe the Frog, Happy Merchant), and topics of discussion using words with racial/hateful meaning (e.g., jewish, monkey, nigger), which may be an indicator that groups adjacent to the alt-right are using genetic testing to bolster their ideology. 

\section{Language Analysis}
Although they both provide discussion platforms, Reddit and 4chan operate in different ways:
e.g., the former requires registration, while the predominant mode of operation on the latter is via anonymous and ephemeral posting.%
\longVer{ Also, Reddit supports an infinite number of user-driven sub-communities, while on 4chan conversations are organized around a few dozens boards, with images playing a key role.}
Naturally, they also attract different sets of users and content, e.g., 4chan is typically identified as a fringe community, while, Reddit, though also hosting fringe communities, is overall a mainstream site (5th most visited in the US).

Our analysis of genetic testing on the two platforms thus far has highlighted that genetic testing is a subject which is discussed frequently; on Reddit, in subreddits ranging many aspects of the every day life of the users, on 4chan, in threads that attract an order of magnitude more posts. %
At the same time, on both platforms, fringe political groups express their wish to marginalize minorities using genetic testing.
Next, we provide a comparison of the {\em language} used in the context of conversations that are likely to include genetic testing.
To do so, we turn to word embeddings, specifically, word2vec~\cite{mikolov2013distributed}.
Word2vec models are trained on large corpora of text, and generate a high-dimensional vector for each word that appears in the corpus;
words that are used in similar context also have a closer mapping to the high-dimensional vector space.
This allows us to study which words are used in similar contexts.

\descr{Methodology.} We train a separate word2vec model, as per the implementation provided by~\cite{rehurek_lrec}, for each of the 19 groups of subreddits (see Figure~\ref{fig:topics_based_on_subreddits}) and 4chan's \dspol, using all of the posts made between January 1, 2016 and March 31, 2018, and June 30, 2016 and March 13, 2018, respectively. %
We pre-process each corpus as follows: 1) we remove special symbols, punctuation, URLs, and numbers; 2) we tokenize each word that appears on each post; and 3) we perform stemming on the words using the Porter algorithm.
Next, we train word2vec models for each community on all the pre-processed posts and all words that appear at least 100 times in each corpus.
We use a {\em context window} equal to 7, i.e., the model considers a context of up to 7 words ahead and behind the current word.

\descr{Vocabulary.} Table~\ref{tbl:w2v_vocabulary_sizes} reports the number of words that are considered in each word2vec model.
Vocabulary sizes vary greatly, e.g., from 122 in the Ancestry subreddits to 46K in Race/Culture subreddits. 
This is due to the fact that we only consider words that appear at least 100 times. %

\begin{table}[t]
\centering
\vspace*{0.2cm}
\footnotesize
\begin{tabular}{@{}lrlr@{}}
\toprule
\textbf{Group} & \multicolumn{1}{l}{\textbf{\begin{tabular}[r]{@{}r@{}}\# of Words \\ in Vocabulary\end{tabular}}} & \textbf{Group} & \multicolumn{1}{l}{\textbf{\begin{tabular}[r]{@{}r@{}}\# of Words\\  in Vocabulary\end{tabular}}} \\ \midrule
4chan's \dspol & 31,337 & Hate & 40,223 \\
Ancestry & 122 & Health & 11,101 \\
Animals & 8,065 & Legal & 4,655 \\
Children & 15,858 & News & 32,097 \\
Crime & 11,649 & Politics & 41,057 \\
Drugs & 7,858 & Race/Countries & 46,978 \\
Educational & 23,151 & Religion & 12,431 \\
Entertainment & 7,743 & Science & 18,341 \\
Funny & 5,641 & Sexes & 20,743 \\
Genetics & 1,178 & Other & 24,767 \\ \bottomrule
\end{tabular}
\vspace{-0.1cm}
\caption{Words that are in the vocabulary of the word2vec models trained for each group of subreddits and \dspol.}
\label{tbl:w2v_vocabulary_sizes}
\vspace{-0.2cm}
\end{table}

\descr{Training.}
To assess how each community discusses topics related to ethnicity and genetic testing words,
for each word2vec model, we get the 10 most similar words for two groups of seed words:
1) 91 genetic testing keywords obtained from the list of 280 keywords (the other 189 including multiple words so we do not consider them)
2) a hand-picked set of words, namely, \enquote{white,} \enquote{black,} \enquote{jew,} \enquote{kike,} \enquote{ancestry,} \enquote{dna,} and \enquote{test.} 
The latter are added aiming to assess whether ethnic terms (e.g., ``white'') and genetic testing keywords (e.g., ``dna'') are used in different contexts than the set of genetic keywords (e.g., ``23andMe''). 

\begin{figure*}[t!]
  \center
  \subfigure[Genetic Testing Keywords]{\includegraphics[width=0.6\columnwidth]{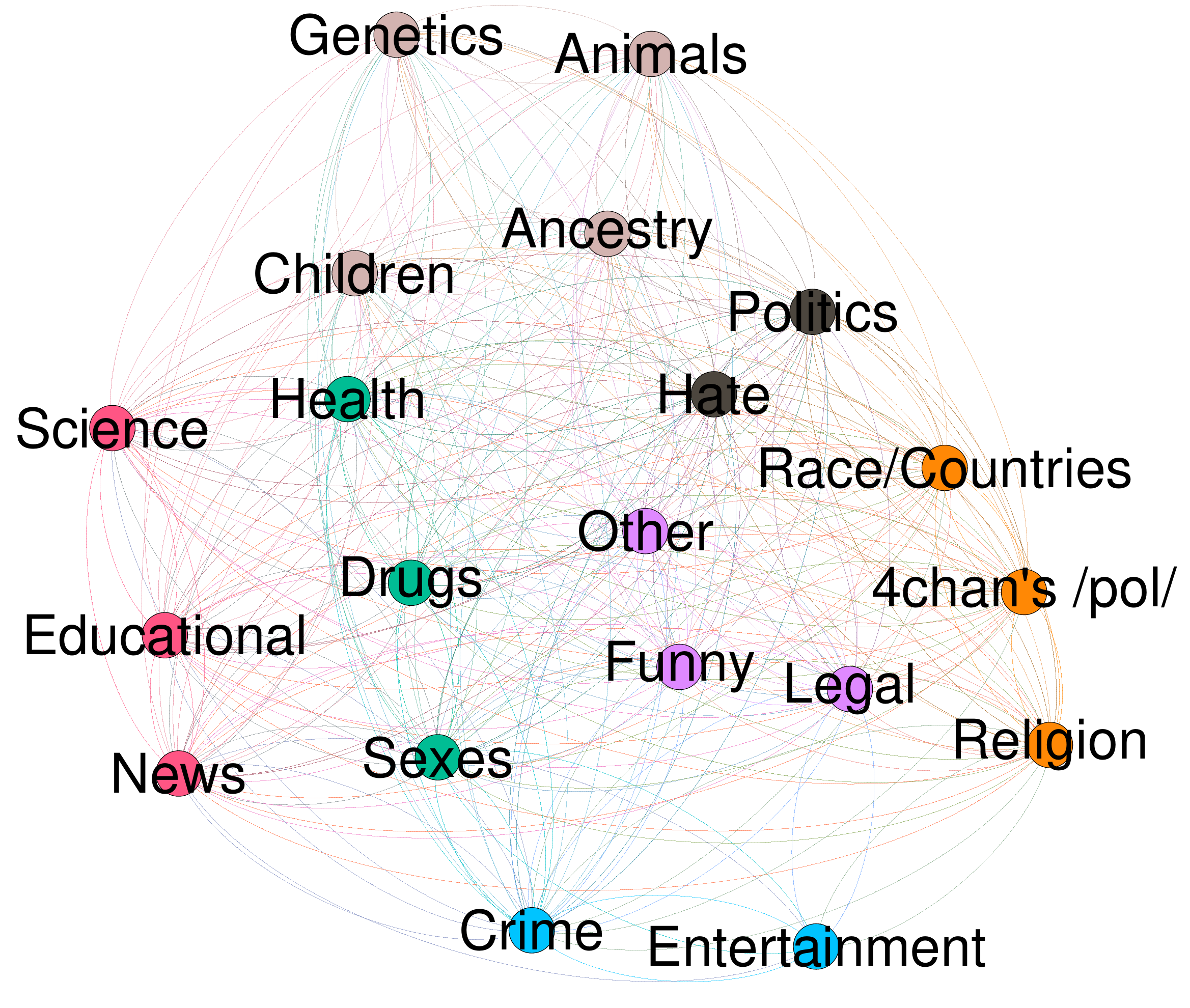}\label{fig:w2v-genetic}}
~~~~~~~~
  \subfigure[Selected Keywords]{\includegraphics[width=0.6\columnwidth]{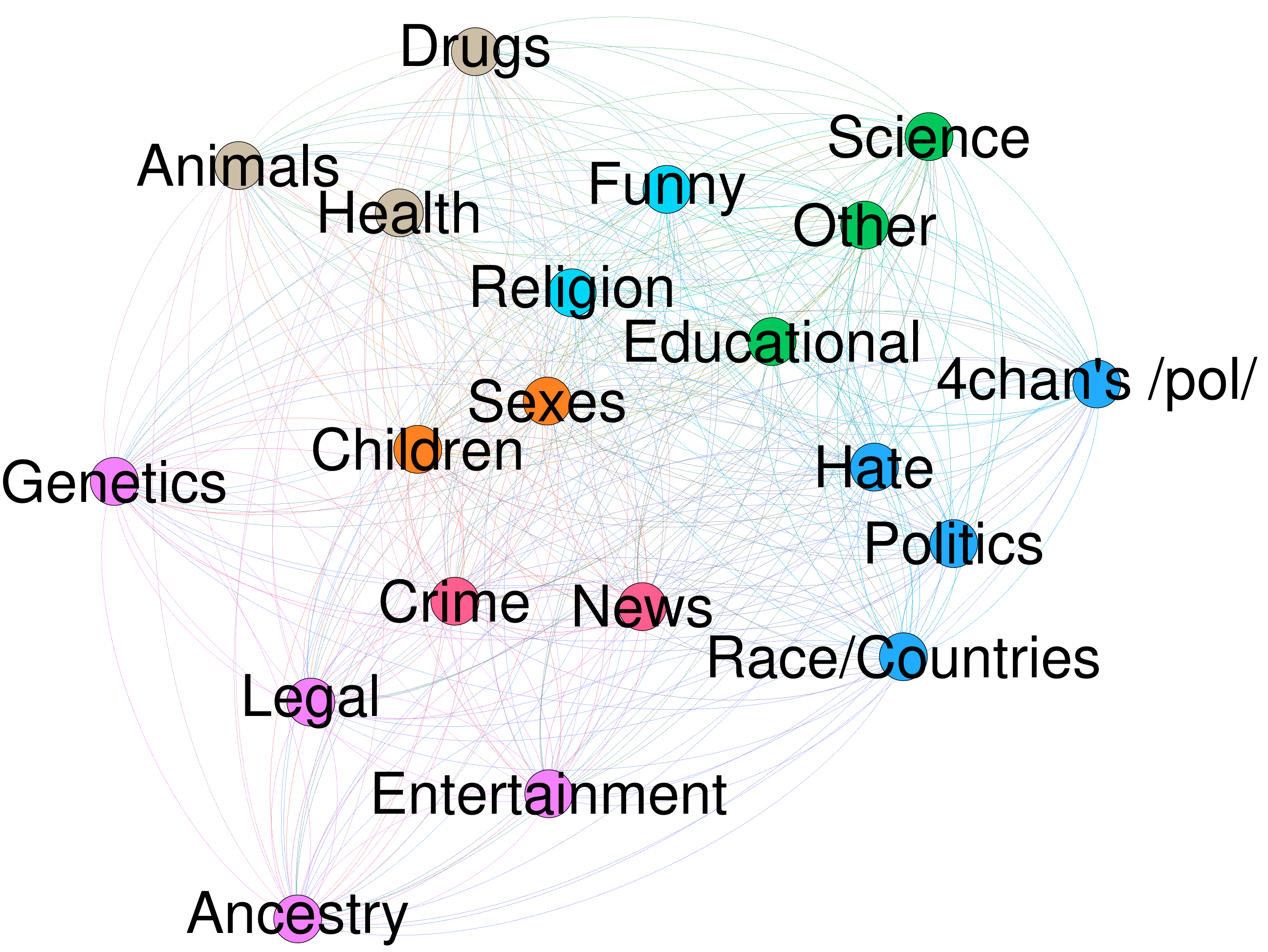}\label{fig:w2v-picked}}
  \caption{Graph representation of the word2vec models, using as seeds: (a) all the genetic testing keywords, (b) the terms ``white,'' ``black,'' ``jew,'' ``kike,'' ``ancestry,'' ``dna,'' and ``test.'' See~\cite{fig7a_interactive,fig7b_interactive} respectively for interactive versions.}
  \label{fig:w2v_models_graph}
\end{figure*}

\descr{Visualization.} We calculate the \emph{similarity} of all the possible combinations of word2vec models using the Jaccard Index scores of all the similar words for all the seed words.
Then, we create two complete graphs (see Figure~\ref{fig:w2v_models_graph}), one for each set of seed keywords, where nodes are the trained word2vec models and edges are weighted by the Jaccard Index score between the similar words for all the seed words.
Once again, we use the community detection algorithm by~\cite{blondel2008fast}. %
\longVer{

By looking at the communities, we gather some interesting insights about how communities compare to each other. }%
When using the genetic testing keywords as seeds (Figure~\ref{fig:w2v-genetic}), we find that communities about genetics, ancestry, animals, and children discuss genetic testing in very similar contexts (light brown nodes). 
Similarly, we find a cluster with subreddits with scientific, educational, and news content (red nodes on the left), and another related to health, drugs, and sexes (green nodes). 
Interestingly, the subreddits in the hate category discuss genetic testing in a similar manner as the political ones (brown nodes); 
this is not entirely surprising also considering that these categories have the two highest toxicity levels (cf.~Figure~\ref{fig:subreddits-perspective}). 
Also, \dspol users seem to discuss genetic testing in a context similar to subreddits related to race/countries and religion (orange nodes). %
This may be because \dspol frequently discusses Judaism (with references to Israel and the Jewish community), as well as other religions~\cite{Finkelstein2018}.

When using the set of hand-picked seed words (Figure~\ref{fig:w2v-picked}), \dspol is similar to the hateful subreddits, as well as the subreddits about politics and race/countries (blue nodes).
In other words, Hate, Politics, Race/Countries subreddits, and \dspol, use ethnic terms in conjunction with genetic testing keywords in similar contexts.
Overall, the fact that that certain subreddits share language characteristics with \dspol is particularly worrying as it may be an indicator of 4chan's fringe ideologies propagating into more mainstream media.

\section{Discussion \& Conclusion}\label{sec:conclusion}

Direct-to-consumer (DTC) genetic testing is one of the first revolutionary technologies with the potential to transform society by improving people's lives.
Citizens of most developed countries have easy and affordable access to a wealth of informative reports, which allow them to better understand themselves, learn about their health and their cultural heritage, and find lost relatives~\cite{borrelli2018relatives}. 
However, this new technology also harbors societal dangers as it is used by fringe groups as ``evidence'' on which to build discrimination and prejudice, and potentially increase ethnic sectarianism.
Considering how information has become increasingly misused on the Web, the potential abuse of genetic testing on online platforms is not be underestimated.

Nevertheless, prior work on this topic has mostly been limited to qualitative studies~\cite{Panofsky2017,Roth2018}, which discuss how DTC genetic testing may have a negative societal impact due to their results possibly reinforcing the concept of racial privilege. 
In that respect, our analysis furthers this line of research by taking a large-scale, data-driven approach, which provides new insight into both the breadth and depth of the issue (of which hate speech is an important aspect). 
We believe that our findings broaden the discussion around DTC genetic testing and its potential misuse in furthering hateful rhetoric and ideology as we provide quantitative evidence for the prior qualitative work. 

More specifically, we shed light on online discussions about genetic testing on two social networks, Reddit and 4chan's politically incorrect board (\dspol), which are known to include fringe and alt-right communities.
We analyzed 1.3M comments spanning 27 months using a set of 280 keywords related to genetic testing as search terms, relying on a mix of tools including Latent Dirichlet Allocation, Google's Perspective API, Perceptual Hashing, and word embeddings to identify trends, themes, and topics of discussion.
Our analysis showed that genetic testing is frequently discussed on both platforms.
For instance, on \dspol, we find an order of magnitude increase in activity on threads related to genetic testing when compared to a random sample. Interestingly, images appearing along genetic testing conversations often include alt-right personalities and antisemitic memes.
On Reddit, genetic testing is discussed in a wider variety of contexts, however, while there are communities building around the more positive aspects (e.g., health, cultural heritage, etc.), we also found others where conversations include racist, hateful, and misogynistic content. 

Overall, we uncovered evidence of genetic testing being misused in online discussions, %
further ingraining and empowering genetics-based prejudice, discrimination, and even calls for genocide.
For instance, comments on both \dspol and a set of ``hateful'' subreddits often contain highly toxic language, with users even suggesting leveraging genetic testing tools to further marginalize or even eliminate minorities.  
In fact, word embeddings showed that \dspol and certain subreddits share worrying language characteristics, which may be an indicator of 4chan's fringe ideologies spilling out to more mainstream platforms. 

Our findings are particularly timely as recent events indicate that those interested in societal disruption have successfully seized upon technological innovations and used them in ways that were not intended by their creators.
More specifically, information has been increasingly weaponized, including by state actors, to sew racial discontent~\cite{stewart2018examining} and even instigate public health crises~\cite{broniatowski2018weaponized}.
In this context, recent efforts have been made by law enforcement to understand and address such campaigns~\cite{muellerreport}. 
Thus, we ought to reflect on the practical implications of our findings and how they affect future work in this area.
Considering that previous qualitative studies~\cite{Panofsky2017,Roth2018} demonstrate how the commercialization of genetic testing may have a negative societal impact, and since our study provides quantitative data on the matter, the next natural step is to examine whether genetic ancestry testing has an (indirect) effect on the levels of racism and discrimination online. 
Naturally, such correlation is not easy to identify and it may require a mixed-methods methodological approach (e.g., interviews with people adjacent to the far-right),
but our work arguably provides a stepping stone toward this.

Finally, we note that platforms like Facebook and Twitter have begun to be held accountable when their services enable harmful behavior~\cite{facebook2019accountable}; 
if there are strong indications that DTC genetic ancestry testing exacerbates online discrimination, we believe that the DTC industry should also consider the potential abuse of their services and attempt to find ways of minimizing this behavior. 
In future work, we plan to build tools that automatically distinguish healthy from toxic comments about genetic testing. 
Currently, a number of techniques (e.g.,~\cite{Davidson2017,del2017hate,gao2017detecting,djuric2015hate,gamback2017using,gao2017recognizing}) 
are available that can be used to identify hateful/toxic comments, using machine learning models trained on annotated datasets. 
We plan to use similar methods on the dataset built in this work to train models that identify toxic comments specifically in the context of genetic testing, confident that this will yield better accuracy than generic ones.

\descr{Acknowledgments.} 
 This project has received funding from the European Union's Horizon 2020 Research and Innovation program under the Marie  Skłodowska-Curie \enquote{Privacy\&Us} and \enquote{ENCASE} projects (Grant Agreement No. 675730 and 691025), as well as a Google Faculty Award. 
We also gratefully acknowledge the support of the NVIDIA Corporation for donating two Titan Xp GPUs used in our experiments.

{\footnotesize

\bibliographystyle{abbrv}
\input{no_comments.bbl}}

\appendix

\section{List of Subreddits}

In Table~\ref{tab:most_common_subreddits}, we report the list of subreddits sorted by normalized number of genetic testing comments.

\input{tables/subreddits}

\end{document}

%% file: tables/subreddits.tex
\begin{table*}[t]
\setlength{\tabcolsep}{4pt}
\resizebox{\textwidth}{!}{%
\begin{tabular}{@{}rlrrrl|rlrrrl@{}}
\toprule
& \textbf{Subreddit}  & \textbf{Gen Test}  & \textbf{Total}         & \textbf{Percent.}        & \textbf{Tag} &  & \textbf{Subreddit}          
& \textbf{Gen Test}   & \textbf{Total}     & \textbf{Percent.}      & \textbf{Tag} \\[-0.4ex]
& & {\bf Comms}       & {\bf Comms}        & & & & & {\bf Comms} & {\bf Comms}\\
\midrule
1            & r/promethease          & 347                                               & 2,580                                             & $13\%$                & Genetics              & 58              & r/tifu                 & 390                                  & 2,191,142                           & 0.01\%                    & Other                                              \\ 
2            & r/SNPedia              & 184                                               & 1,774                                             & $10\%$                & Genetics              & 59              & r/TwoXChromosomes      & 488                                  & 2,753,369                           & 0.01\%                    & Sexes                                              \\                     
3            & r/23andme              & 4,150                                             & 44,225                                            & $9\%$                 & Genetics              & 60              & r/breakingmom          & 101                                  & 609,366                             & 0.01\%                    & Children                                           \\                                 
4            & r/Ancestry             & 190                                               & 2,793                                             & $6\%$                 & Ancestry              & 61              & r/Advice               & 157                                  & 1,021,798                           & 0.01\%                    & Other                                              \\ 
5            & r/Genealogy            & 3569                                              & 95,205                                            & $3\%$                 & Ancestry              & 62              & r/PurplePillDebate     & 210                                  & 1,421,805                           & 0.01\%                    & Hate                                               \\
6            & r/genetics             & 347                                               & 11,741                                            & $2\%$                 & Genetics              & 63              & r/aww                  & 799                                  & 5,671,423                           & 0.01\%                    & Other                                              \\ 
7            & r/Adoption             & 610                                               & 40,667                                            & $1\%$                 & Children              & 64              & r/history              & 142                                  & 1,054,177                           & 0.01\%                    & Other                                              \\
8            & r/IDmydog              & 175                                               & 14,429                                            & $1\%$                 & Animals               & 65              & r/raisedbynarcissists  & 163                                  & 1,214,553                           & 0.01\%                    & Other                                              \\
9            & r/ehlersdanlos         & 340                                               & 47,303                                            & $0.7\%$               & Health                & 66              & r/milliondollarextreme & 109                                  & 895,032                             & 0.01\%                    & Hate                                               \\
10           & r/TheBlackList         & 288                                               & 43,127                                            & $0.6\%$               & Entertainment         & 67              & r/asktransgender       & 158                                  & 1,307,753                           & 0.01\%                    & Sexes                                              \\
11           & r/Celiac               & 171                                               & 41,444                                            & $0.4\%$               & Health                & 68              & r/exmormon             & 288                                  & 2,444,535                           & 0.01\%                    & Religion                                           \\ 
12           & r/Testosterone         & 306                                               & 83,997                                            & $0.3\%$               & Health                & 69              & r/nottheonion          & 313                                  & 2,898,542                           & 0.01\%                    & News                                               \\
13           & r/serialpodcast        & 745                                               & 213,958                                           & $0.3\%$               & Entertainment         & 70              & r/MapPorn              & 114                                  & 1,063,518                           & 0.01\%                    & Other                                              \\
14           & r/EARONS               & 155                                               & 48,613                                            & $0.3\%$               & Crime                 & 71              & r/explainlikeimfive    & 388                                  & 3,741,174                           & 0.01\%                    & Educational                                        \\
15           & r/StevenAveryIsGuilty  & 357                                               & 126,689                                           & $0.2\%$               & Crime                 & 72              & r/Futurology           & 278                                  & 2,689,784                           & 0.01\%                    & Science                                            \\
16           & r/cancer               & 172                                               & 68,037                                            & $0.2\%$               & Health                & 73              & r/NoStupidQuestions    & 198                                  & 1,943,855                           & 0.01\%                    & Educational                                        \\
17           & r/dogs                 & 1,627                                             & 803,094                                           & $0.2\%$               & Animals               & 74              & r/AskWomen             & 324                                  & 3,328,046                           & ${<}0.01\%$                   & Sexes                                              \\
18           & r/MakingaMurderer      & 1,198                                             & 624,641                                           & $0.1\%$               & Crime                 & 75              & r/UpliftingNews        & 114                                  & 1,214,761                           & ${<}0.01\%$                   & News                                               \\
19           & r/SuperMaM             & 139                                               & 73,997                                            & $0.1\%$               & Crime                 & 76              & r/Documentaries        & 130                                  & 1,386,157                           & ${<}0.01\%$                   & Educational                                        \\ 
20           & r/Nootropics           & 613                                               & 331,434                                           & $0.1\%$               & Drugs                 & 77              & r/todayilearned        & 1,185                                & 13,088,194                          & ${<}0.01\%$                   & Educational                                        \\
21           & r/DebateAltRight       & 298                                               & 169,354                                           & $0.1\%$               & Hate                  & 78              & r/conspiracy           & 469                                  & 5,281,831                           & ${<}0.01\%$                   & Other                                              \\
22           & r/AugustBumpers2017    & 120                                               & 71,825                                            & $0.1\%$               & Children              & 79              & r/news                 & 1,717                                & 19,386,087                          & ${<}0.01\%$                   & News                                               \\
23           & r/ttcafterloss         & 223                                               & 141,992                                           & $0.1\%$               & Children              & 80              & r/ireland              & 138                                  & 1,615,105                           & ${<}0.01\%$                   & Race/Countries                                     \\
24           & r/UnresolvedMysteries  & 966                                               & 667,940                                           & $0.1\%$               & Crime                 & 81              & r/TumblrInAction       & 216                                  & 2,563,058                           & ${<}0.01\%$                   & Hate                                               \\
25           & r/InfertilityBabies    & 156                                               & 111,862                                           & $0.1\%$               & Children              & 82              & r/depression           & 103                                  & 1,277,435                           & ${<}0.01\%$                   & Health                                             \\
26           & r/BeforeNAfterAdoption & 107                                               & 81,078                                            & $0.1\%$               & Animals               & 83              & r/askscience           & 101                                  & 1,289,247                           & ${<}0.01\%$                   & Science                                            \\
27           & r/TickTockManitowoc    & 443                                               & 364,725                                           & $0.1\%$               & Crime                 & 84              & r/fatlogic             & 120                                  & 1,543,070                           & ${<}0.01\%$                   & Hate                                               \\
28           & r/pitbulls             & 108                                               & 103,844                                           & $0.1\%$               & Animals               & 85              & r/IAmA                 & 242                                  & 3,521,706                           & ${<}0.01\%$                   & Other                                              \\
29           & r/infertility          & 427                                               & 423,863                                           & $0.1\%$               & Children              & 86              & r/technology           & 268                                  & 4,072,195                           & ${<}0.01\%$                   & Technology                                         \\
30           & r/arabs                & 128                                               & 157,054                                           & $0.08\%$              & Race/Countries        & 87              & r/AdviceAnimals        & 372                                  & 5,906,232                           & ${<}0.01\%$                   & Other                                              \\
31           & r/BabyBumps            & 973                                               & 130,1608                                          & $0.07\%$              & Children              & 88              & r/Showerthoughts       & 477                                  & 8,034,239                           & ${<}0.01\%$                   & Other                                              \\
32           & r/altright             & 108                                               & 166,436                                           & $0.06\%$              & Hate                  & 89              & r/trashy               & 110                                  & 1,897,268                           & ${<}0.01\%$                   & Funny                                              \\
33           & r/Judaism              & 178                                               & 299,667                                           & $0.06\%$              & Race/Countries        & 90              & r/BlackPeopleTwitter   & 209                                  & 3,762,278                           & ${<}0.01\%$                   & Hate                                               \\
34           & r/AskDocs              & 193                                               & 385,831                                           & $0.05\%$              & Health                & 91              & r/OldSchoolCool        & 142                                  & 2,593,419                           & ${<}0.01\%$                   & Other                                              \\
35           & r/TryingForABaby       & 192                                               & 411,263                                           & $0.04\%$              & Children              & 92              & r/canada               & 231                                  & 4,341,997                           & ${<}0.01\%$                   & Race/Countries                                     \\
36           & r/slatestarcodex       & 123                                               & 273,357                                           & $0.04\%$              & Science               & 93              & r/CringeAnarchy        & 217                                  & 4,101,269                           & ${<}0.01\%$                   & Politics                                           \\
37           & r/bipolar              & 164                                               & 396,899                                           & $0.04\%$              & Health                & 94              & r/AskMen               & 195                                  & 3,805,036                           & ${<}0.01\%$                   & Sexes                                              \\
38           & r/MensRights           & 399                                               & 993,039                                           & $0.04\%$              & Sexes                 & 95              & r/The\_Donald          & 1,251                                & 28,360,073                          & ${<}0.01\%$                   & Hate                                               \\
39           & r/bestoflegaladvice    & 144                                               & 362,868                                           & $0.03\%$              & Legal                 & 96              & r/worldnews            & 845                                  & 20,224,373                          & ${<}0.01\%$                   & News                                               \\
40           & r/steroids             & 320                                               & 825,647                                           & $0.03\%$              & Drugs                 & 97              & r/europe               & 219                                  & 5,275,810                           & ${<}0.01\%$                   & Race/Countries                                     \\ 
41           & r/legaladvice          & 1,081                                             & 2,851,210                                         & $0.03\%$              & Legal                 & 98              & r/atheism              & 108                                  & 2,626,435                           & ${<}0.01\%$                   & Religion                                           \\
42           & r/hapas                & 128                                               & 368,467                                           & $0.03\%$              & Race/Countries        & 99              & r/AskReddit            & 5,421                                & 132,899,306                         & ${<}0.01\%$                   & Other                                              \\
43           & r/science              & 782                                               & 2,666,213                                         & $0.03\%$              & Science               & 100             & r/india                & 127                                  & 3,141,858                           & ${<}0.01\%$                   & Race/Countries                                     \\
44           & r/ADHD                 & 168                                               & 576,203                                           & $0.03\%$              & Health                & 101             & r/KotakuInAction       & 109                                  & 2,811,180                           & ${<}0.01\%$                   & Hate                                               \\
45           & r/changemyview         & 538                                               & 1,908,120                                         & $0.02\%$              & Other                 & 102             & r/pics                 & 543                                  & 15,528,294                          & ${<}0.01\%$                   & Other                                              \\
46           & r/TheRedPill           & 270                                               & 1,044,079                                         & $0.02\%$              & Hate                  & 103             & r/politics             & 1,517                                & 46,270,193                          & ${<}0.01\%$                   & Politics                                           \\
47           & r/confession           & 182                                               & 710,132                                           & $0.02\%$              & Other                 & 104             & r/personalfinance      & 150                                  & 4,671,327                           & ${<}0.01\%$                   & Other                                              \\
48           & r/teenmom              & 203                                               & 824,312                                           & $0.02\%$              & Entertainment         & 105             & r/Philippines          & 102                                  & 3,245,641                           & ${<}0.01\%$                   & Race/Countries                                     \\
49           & r/TeenMomOGandTeenMom2 & 133                                               & 565,612                                           & $0.02\%$              & Entertainment         & 106             & r/unitedkingdom        & 105                                  & 3,595,982                           & ${<}0.01\%$                   & Race/Countries                                     \\
50           & r/Parenting            & 194                                               & 829,177                                           & $0.02\%$              & Children              & 107             & r/ukpolitics           & 125                                  & 4,348,955                           & ${<}0.01\%$                   & Race/Countries                                     \\
51           & r/childfree            & 350                                               & 1,531,152                                         & $0.02\%$              & Children              & 108             & r/trees                & 113                                  & 4,009,217                           & ${<}0.01\%$                   & Drugs                                              \\
52           & r/MGTOW                & 365                                               & 1,625,881                                         & $0.02\%$              & Hate                  & 109             & r/WTF                  & 131                                  & 5,609,346                           & ${<}0.01\%$                   & Other                                              \\
53           & r/relationship\_advice & 309                                               & 1,383,111                                         & $0.02\%$              & Other                 & 110             & r/videos               & 319                                  & 13,934,560                          & ${<}0.01\%$                   & Other                                              \\
54           & r/ShitAmericansSay     & 122                                               & 547,506                                           & $0.02\%$              & Comedy                & 111             & r/funny                & 321                                  & 15,792,122                          & ${<}0.01\%$                   & Funny                                              \\
55           & r/relationships        & 1,853                                             & 8,538,031                                         & $0.02\%$              & Other                 & 112             & r/gifs                 & 111                                  & 9,032,723                           & ${<}0.01\%$                   & Other                                              \\
56           & r/JUSTNOMIL            & 359                                               & 1,790,725                                         & $0.02\%$              & Other                 & 113             & r/movies               & 111                                  & 11,810,334                          & ${<}0.01\%$                   & Other                                              \\
57           & r/TrueReddit           & 102                                               & 557,598                                           & $0.01\%$              & Other                 & 114             & r/nba                  & 183                                  & 23,109,676                          & ${<}0.01\%$                   & Other                                              \\ \bottomrule
\end{tabular}%
}
\caption{List of subreddits sorted by normalized number of genetic testing comments.}
\label{tab:most_common_subreddits}
\vspace{-0.5cm}
\end{table*}